Review paper

# The Epigenetic Tapestry:
## A Review of DNA Methylation and Non-Coding RNA's Interplay with Genetic Threads, Weaving a Network Impacting Gene Expression and Disease Manifestations

Yu-Li He[1*], Youshin Loh[1]

[1] International Bilingual School at Hsinchu Science Park Jieshou Rd. 300, East District, Hsinchu City, Republic of China (Taiwan)
*Corresponding author: heyuli2008@gmail.com

## ABSTRACT
The emerging field of epigenetics has recently unveiled a dynamic landscape in which gene expression is not determined solely by genetic sequences but also by intricate regulatory mechanisms. This review examines the interactions between these regulatory mechanisms, including DNA methylation and non-coding RNAs (ncRNAs), that orchestrate gene expression fine-tuning for cellular homeostasis and the pathogenesis of a multitude of diseases. We explore long non-coding RNAs (lncRNAs) such as telomeric repeat-containing RNA (TERRA) and Fendrr, highlighting their role in protein regulation to ensure proper gene activation or silencing. Additionally, we explain the therapeutic potential of brain-derived neurotrophic factor (BDNF)-related microRNA 132, which has shown promise in treating chronic illnesses by restoring BDNF levels. Finally, this review covers the role of DNA methyltransferases and ncRNAs in cancer, focusing on how lncRNAs contribute to X chromosome inactivation and interact with chromatin-modifying complexes and DNA methyltransferase inhibitors to reduce cancer cell aggressiveness. By amalgamating the wide array of research in this field, we aim to provide glimpses into the complex entangling of genetics and environment as they control gene expressions.

## INTRODUCTION
Epigenetics is an emerging area of genetics research, playing an important role in developing humans and their respective diseases. Most epigenetic effects are characterized by their ability to be reversed, or modified further given the appropriate conditions. This paper discusses the intricate interplay between epigenetics and ncRNA revealing a complex regulatory network that plays a role in the fine-tuning of gene expression, impacting cellular homeostasis and disease development. By relating epigenetics with ncRNA one can predict and influence the development and progression of adverse epigenetic inheritance such as obesity, type 2 diabetes, and mental health disorders, including schizophrenia, in situations such as the Dutch Hunger Winter (Kaspar et al., 2020). Understanding how ncRNAs serve as intermediaries between the epigenetic marks and the actual gene expression allows us to identify them and use them as biomarkers for disease risk assessment. We present several examples of epigenetic changes causing aberrant expression of ncRNAs in the context of tumor progression. Next, as ncRNAs can also be used as biomarkers for diagnosis and prognosis or explored as potential targets, we present insights into the use of ncRNAs for targeted cancer therapy. (Ferreira & Esteller, 2018)

Classic examples of epigenetic results in humans include variations in height. While one may have the genetics coding for a tall height, environmental factors, including nutrition and contagious diseases, may limit that development. Thus emotional, mental, and physical surroundings like high stress and depression-causing conditions can result in the discordance of the physical characteristics of identical twins. These "epi-factors" may cause gene expression, neural circuit function, and behavior changes. These epi-inheritances also vary depending on when the environment is contacted, whether it is in the individual's developmental stages or adult stages. Nevertheless, discrepancies in one's surroundings are shown to initiate modifications in specific brain regions (Nestler et al., 2016).

Questions such as phenotypic variations between twins may have arisen from ncRNA's effects on genes that code for important matters such as disease vulnerability and even appearance. Specific ncRNAs such as lncRNAs are already known to affect epigenetic characteristics through histone modification, post-transcriptional regulation, and DNA methylation. They play a crucial role in the regulatory network of gene expression controlled by changing the states of chromatin, either in euchromatin, a state ready for gene expression, or heterochromatin, a state not fit for expression. This is only one example of the many different ways in which ncRNAs and epigenetics are related. They are interconnected in many more ways and can be used for better understanding and preventing diseases. The same effect can be observed not through ncRNAs, but through enzymes named DNA methyltransferases, such as those in bacteria, which have three major forms of DNA methylation, some in common with eukaryotes: 5-methylcytosine (m5C), N6-methyladenosine (m6A), and N4-methylcytosine (m4C). DNA methyltransferase (DNMTs) adds methyl groups to specific DNA locations, such as the C5 or N4 position of cytosine and the N6 position of adenine, these DNA methyltransferases have the opportunity to interact with ncRNAs to exhibit the effects mentioned. (Wang et al., 2023)

Recently many new studies have shown the importance of ncRNAs in epigenetic processes. The intricate processes between ncRNAs and epigenetics are still not fully discovered but the identification of more and more ncRNAs is slowly unveiling the tapestry woven by their interactions. While epigenetics is certainly a factual scientific explanation, there are some misconceptions in the general population and exceptions to the theory. For example, many hope that playing classical music to newborns can enhance cognitive development through epigenetic means, but there is no scientific evidence yet to support this claim. The real factor in enhancing cognitive skills is hands-on learning and memory development that causes released repeated activation of brain circuits. ncDNAs are simply sequences of DNA that do not code for the production of a protein, but they are not to be confused with "junk", a term coined by Susumu Ohno in 1972. His term has received backlash from many scholars citing that through evolution, "junk" DNA would have been evolved out, but given that they are still present nearly ubiquitously in various organisms, it must serve a function in maintaining the homeostasis of the individual that evolutionarily keeps its "junk" ncRNAs. Small interfering RNA(siRNA), micro RNA (miRNA), piwi-interacting RNA (piRNA), and ncRNA are some of the more well-known ncRNAs with imprintation functions in gene regulation caused by epigenetics.

To specify one example, lncRNA is typically more than 200 nucleotides in length and is found in the nucleus or cytoplasm. In recent studies, scientists have found that lncRNAs play a role in epigenetic regulation for genetic imprinting and X chromosome inactivation. In the case of X chromosome inactivation, one type of lncRNA, Xist RNA, stays in the nucleus and interacts physically with the X chromosome that is about to be inactivated by Xist RNA. (Zhao et al., 2008b) showed that a fragment of Xist RNA can induce the 27th amino acid in histone (H3H3K27) trimethylation, a process of fear imprintation to the genes in rats, by recruiting the Polycomb repressive complex 2 (PRC2), a protein complex responsible for the regulation of many genes in H3K27, to the inactivated X [(Handbook of Clinical Neurology | Elsevier, n.d.)](#). Furthermore, allosteric regulation is directly involved with lncRNA and its ability to regulate epigenetics.

The regulation of lncRNAs occurs through the binding of proteins to one of the many sites in lncRNA, resulting in a change in shape which will affect the expression of genes initiated by the altered lncRNA. lncRNAs usually have a larger number of these binding sites compared to other RNAs indicating the importance of allostery in epigenetics. [(C. Wang et al., 2017)](#) This allows us to understand the processes involved in epigenetics about lncRNA, as well as the concepts that another ncRNA utilizes.

Through the analysis and presentations of the multi-fielded studies, some of the top academic papers are broken down into easily understandable segments, allowing readers of all levels of expertise to thoroughly understand the underlying mechanism by which epigenetics takes place. This paper will explore epigenetics and non-coding genetic material in several sections with different focuses, such as a focus on the mechanisms of epigenetic traits using lncRNAs and DNA methyltransferases with their corresponding inhibitors, and a focus on case studies of epigenetic effects, with topics including epi-transcriptomic activities, plant epigenetics with an exploration of inheritable epigenetics, chemoresistance, chronic pain regulation, major depressive disorder treatments, and a specific RNA called TERRA.

## ncRNA

ncRNAs, which have been established to be major players in epigenetics, act epigenetically on an extremely ubiquitous enzyme. Its interactions with ncRNAs are discussed. The intricate interactions between RNA polymerase II, ncRNAs, and the regulatory mechanisms that control gene expression have been pioneered by a few leading papers in the field. One such article by Espinoza et al. sheds light on the obscure regulatory mechanisms controlling gene expression by exploring the complex mechanisms of interactions between RNA polymerase II and ncRNA. [(Espinoza et al., 2007)](#) Through the use of fluorescently labeled double-stranded DNA fragments containing the AdMLP region in place of plasmid DNA, the work thoroughly investigates complex formation via native PAGE and clarifies the elongation complex building process. The research reveals the structural and functional nuances of B2 RNA, emphasizing its role in inhibiting mRNA transcription, using painstaking in vitro transcription experiments and thorough analysis of RNA structures utilizing RNase probing, in-line probing, and RNase footprinting. The explanation of how B2 RNA suppresses transcription by attaching to RNA polymerase II's transcription bubble and stopping the enzyme from generating mRNA is one particular example from the study. Through this interaction, ncRNA directly affects gene expression without changing the DNA sequence, illustrating an epigenetic mechanism. B2 RNA functions as a transcriptional repressor by binding to RNA polymerase II, demonstrating how ncRNAs can influence the control of gene expression through epigenetic mechanisms. The study also demonstrated a clear connection between B2 RNA and the field of epigenetics by showing that the inhibitory effect and binding affinity of B2 RNA are dependent on the presence of particular transcription factors and chromatin states. For instance, B2 RNA's regulatory function is context-dependent and influenced by the cell's epigenetic landscape, as evidenced by chromatin immunoprecipitation (ChIP) experiments showing that it binds with restrictive histone modifications like H3K27me3. [(Espinoza et al., 2007b)](#) ncRNAs have a myriad of effects on disease manifestations, of which those like cancer, major depressive disorder, homeostasis & metabolism, and chronic pain will be discussed. In oncological epigenetics, the general mechanism consists of the inhibition of unregulated methylation of tumor suppressor genes, which allows the cancer to grow uncontrollably. In major depressive disorder, the levels of brain-derived neurotrophic factors have been imbalanced, and through the use of certain drugs causing epigenetic change the levels of the brain-derived neurotrophic factors approach that of healthy controls. Similar lncRNA and ncRNA effects have been shown to cause disruptions to the way that neurons and other hormones function such that the addition of certain epigenetic factors will improve a person's chronic pain and homeostasis.

# lncRNA

Epigenetic regulation is a complex process involving various intricate mechanisms that all work together to form the complex tapestry central to gene regulation. An example of this complexity is provided by a study performed by Tsai et al. that examines the regulation of DNA G-quadruplex (G4) structures and ATRX (Alpha Thalassemia/Mental Retardation Syndrome X-linked) occupancy to modify gene expression by TERRA (Telomeric Repeat-containing RNA), a well-known noncoding RNA (Tsai et al., 2022). This study shows the interaction between ncRNAs, epigenetic regulation, and gene expression by showing how the presence of G4 structures, especially in the first exon-intron region of TERRA is negatively affected by the decreased expression of TERRA. According to the findings of the study, TERRA is central in maintaining the integrity of a number of molecules known to impact gene regulation which include the G4 structures. synthesis has been associated with G4 structures and replication, transcription, and genome stability. The work shows that G4 structures at TSS regions decrease upon TERRA knockdown, especially in the 5'-end first exon-intron region. There is evidence that such an area is responsible for the onset of transcription and since G4 structures were found to be located here it can be suggested that these structures play the role of the regulator preventing or allowing the access of the transcription machinery to DNA. In the current study, the authors found that when TERRA is depleted, well-known genes such as MYC and VEGFA, exhibit reduced G4 development in the TSS. This decrease is related to different patterns of gene expression, which demonstrated that TERRA influences G4 structure and gene activity. Regulation of such structures also ensures that, through TERRA, chromatin is set for proper regulation of gene expression. The study also highlights how TERRA impacts ATRX occupancy, a chromatin remodeling protein that recognizes and fixes G4s. ATRX is required to maintain genome stability as well as to regulate the expression of certain genes correctly. It has been shown that under the condition of TERRA depletion, the typical localization of the ATRX protein, namely at the G4 sites of the TSS of the essential cellular genes, gets affected. For example, tumor suppressor genes may be downregulated because of reduced binding of ATRX to the G4 sites near their transcription start site, the function of which may help in carcinogenesis. (Tsai et al., 2022)

The article investigated the potential effects of TERRA depletion on G4 structures since it has been discovered that ATRX plays a role in DNA G4 formation in vivo. Using the BG4 antibody, they carried out immunostaining for DNA G4 structures. Cells were treated with CX-5461, a DNA G4-stabilization ligand, to promote G4 formation to test the sensitivity of BG4 labeling. To remove RNA signals, RNase A was used before DNA G4 immunostaining. When cells were treated with CX-5461, DNA G4 immunostaining showed a substantial increase in DNA G4 signals. Surprisingly, the DNA G4 intensity in TERRA knockdown cells was significantly lower than in control cells, indicating that TERRA maintains the development of DNA G4. (Tsai et al., 2022b)

In the context of lncRNA interactions with proteins, various articles describe the intricate regulation of lncRNA along with the commonly modified nucleosides(nucleotide that lacks a phosphate group) in lncRNA: 6-methyladenosine(m6A), 5-methylcytidine(m5A), pseudouridine(Ψ), and inosine(I). (Kazimierczyk & Wrzesinski, 2021)(Jacob et al., 2017)(Shafik et al., 2016b)

The alteration of RNA metabolism is controlled by three types of proteins. The first category is known as "writers," and they insert changed nucleotides into RNA during the post-transcriptional phase. One example of a writer described in the article is the methylase complex which is connected to the formation of 6-methyl adenosine(m6A) modification. The second group of proteins is called "readers" and they interact with the modified nucleotides introduced by the writers. The last group is involved in the removal of modification labels and is termed "erasers"(Kazimierczyk & Wrzesinski, 2021d).

lncRNAs have also been found to be important players in the epigenetic regulation of gene expression. They affect chromatin state and gene activity in a variety of ways. The development of diseases and many cell functions are profoundly affected by the intricate and two-way relationship between DNA methylation and ncRNAs. In this regard, a study carried out by Luisa Statello et al. demonstrates how lncRNAs help and guide proteins that alter the chromatin structure. (Statello et al., 2020) For instance, long non-coding RNA Fendrr could intersect with Polycomb Repressive Complex 2 (PRC2) and Trithorax Group (TrxG), which trigger deposition of active H3K4me3 marks or repressive H3K27me3 marks respectively (Mangiavacchi et al., 2023). From these attributes, it is clear that lncRNAs specialize in maintaining a delicate homeostasis equilibrate of gene expression necessary for developmental as well as cellular differentiation.

Due to the developmental nature of the organism, Fendrr is able to fine-tune the activity of TrxG and PRC2 to ensure that the right genes are either active or inactive. That is precisely the kind of regulation necessary for the development of structures such as caudal lateral plate mesoderm and overall body morphogenesis.

Additionally, the article explains how RNA-DNA triplex structures allow ncRNAs to interact with chromatin. MEG3, for instance, can create these structures to help PRC2 bind to particular genomic locations, improving H3K27me3 deposition and suppressing gene expression. This method is essential for preserving gene repression, and several malignancies have been associated with MEG3 downregulation, which is frequently caused by promoter hypermethylation. (Mangiavacchi et al., 2023b) Some ncRNAs can directly regulate the expression of genes through their action on chromatin accessibility, which can be illustrated by the example of MEG3 and PRC2. This interaction can be taken as an example of a broad area of lncRNAs that have been shown to act as molecular scaffolds to integrate multiple components of complex regulatory circuits with target chromosomal regions.

This corresponds with the processes that Arianna Mangiavacchi et. al look at in their work. al which elaborates on how lncRNA contributes to epigenetic changes simultaneously. The overview reviews various mechanisms whereby lncRNAs could be involved in gene regulation as enhancers, scaffolds, decoys, or guides. One of the widely renowned examples is Xist, which remains crucial for X-chromosome inactivation due to the demolition of this chromosome since it recruits chromatin modifiers to the X chromosome (Statello et al., 2020b). Transcriptionally condensation is further facilitated by binding of Xist to several proteins such as SHARP and SMRT that inhibit further binding of the RNA polymerase II to the X chromosome. It is believed that PRC1 and PRC2 function in turn to maintain this silent state and that Xist facilitates the recruitment of these complexes by interacting with proteins like hnRNPK (Mangiavacchi et al., 2023c).

Moreover, the processes referring to the regulation of telomeres' length and integrity comprise one of gene control networks based on molecules such as ncRNAs and DNA methylation. To maintain heterochromatin at the telomeres TERRA binds with chromatin modifiers such as TIP5. The ability of TERRA to form RNA-DNA hybrids or R-loops at telomeres indicates its function in the homology-directed repair mechanisms that modulate telomeres that are important in preventing early cellular senescence and ensuring stem cell proliferation(Tsai et al., 2022c). Thus, for the proper cellular homeostasis and target gene expression one should consider an extremely complex and tightly regulated interdependent system which is DNA methylation and ncRNA. The level of regulation achieved by epigenetic mechanisms is evident in the manner in which Fendrr, MEG3, Xist, and TERRA for that of long noncoding RNAs regulate chromatin and gene expression. Comprehending this interaction yields a significant understanding of the mechanisms that underlie epigenetic control and presents prospective targets for therapeutic interventions in diseases like cancer and age-related disorders that are linked to epigenetic dysregulation.

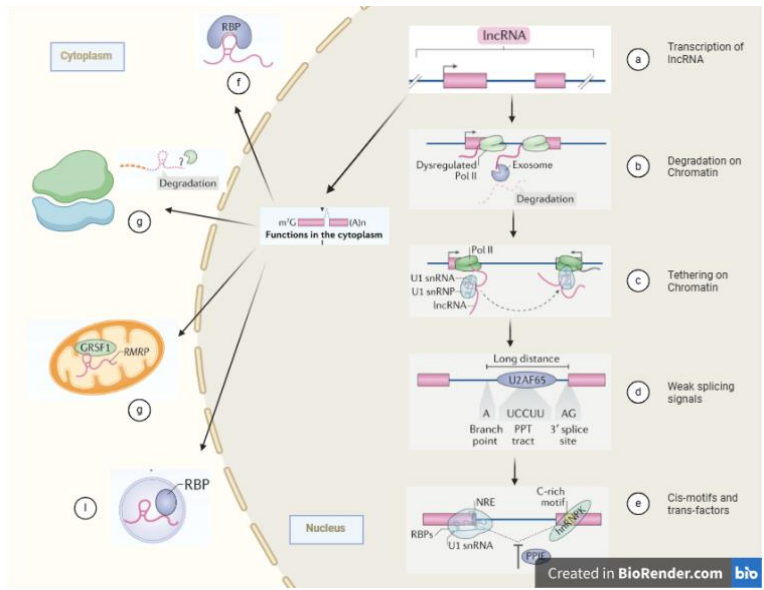

*Figure #1: contrasts the processing of mRNA with the many processes that are involved in the production of lncRNAs. RNA polymerase II (Pol II) transcribes a large number of lncRNAs that are processed inefficiently and are primarily kept in the nucleus. Their nuclear localization is influenced by several mechanisms, including interactions with heterogeneous nuclear ribonucleoproteins (hnRNPs) such as hnRNPK, recruitment of U1 small nuclear ribonucleoprotein (U1 snRNP), and nuclear retention elements (NREs) (Figure description, parts b-e). On the other hand, certain lncRNAs, like those with shorter polypyrimidine tracts (PPTs), frequently experience ineffective splicing, which increases their nuclear retention (Figure description, part d). After entering the cytoplasm, lncRNAs interact with different RBPs, possibly producing complexes that influence the stability and functionality of RBPs. Furthermore, different localization patterns of several lncRNAs, such as exosomal and mitochondrial sorting, indicate that they play other regulatory roles in addition to chromatin modulation. Our knowledge of the intricate and varied roles that lncRNAs play in cellular homeostasis and epigenetic control is enhanced by this thorough understanding of lncRNA biosynthesis.*
*Created with [BioRender.com](BioRender.com)*

      The roles of lncRNA in gene regulation make it an exceptionally favorable biomarker in diseases such as cancer that are usually hard to detect. One such disease, Neuroblastoma (NB) is a type of cancer that affects children under the age of 5. NB arises from immature nerve cells from the adrenal gland, nerve ganglia, or the neck. This article describes the possible use of lncRNA signatures correlated to the survival rate of patients with NB to be utilized for an overall survival time estimator called SVR-NB. SVR stands for "Support Vector Regression" and it is based on support vector regression (SVR)52 and an inheritable bi-objective combinatorial genetic algorithm (IBC GA) (Sathipati et al., 2019). SVR-NB can be used to identify specific lncRNA among a large number of lncRNAs allowing the potential use of lncRNAs as a biomarker. Protein biomarkers have already been extensively searched for to improve NB therapies with certain proteins such as MYCN being expressed in children with Neuroblastoma. Despite these advancements, the identification of targets that are associated with NB is still urgently needed as the long-term survival of the group classified as "High risk" has not considerably improved.

      Various types of ncRNAs, specifically lncRNAs, could act as biomarkers to detect Neuroblastoma early on to improve chances of survival. Various studies have been conducted that indicate the role that lncRNAs play in the development of cancer. NcRNAs have oncogenetic properties and their overexpression is correlated to a higher risk in Neuroblastoma prognosis. RNA sequencing data and survival information from the database of gene expression omnibus (GEO) were used by Srinivasulu Yerukala Sathipati, Divya Sahu, Hsuan-Cheng Huang, Yenching Lin, and Shinn-Ying Ho conclude that "SVR-NB identified 35 out of 783 lncRNAs which are strongly correlated with overall survival in NB patients. SVR-NB using 10-fold cross-validation (10-CV) achieved a mean squared correlation coefficient of 0.85 ± 0.009 and a mean absolute error of 0.56 ± 0.09 years between actual and estimated overall survival times in NB patients" (Sathipati et al., 2019b).

# DNA METHYLATION

Apart from lncRNAs causing epigenetic change, another common form of epigenetic expression is through DNA methyltransferases (DNMTs). One process of DNA methylation works depending on DNMTs (including DNMT1, DNMT3A, DNMT3B, or DNMT3L (Jin & Robertson, 2012) to transfer the methyl groups of S-adenosylmethionine (SAM, a universal methyl group donor molecule) onto the sugar backbone of cytosine in CG, C*G, and C** sequences (where * could be A, T, or C) for eukaryotes and certain non-bacterial organisms. In bacteria, thorough research has determined $m^5C$methyltransferase, $m^6A$ methyltransferase, and $m^4C$ methyltransferase to be the main DNA methyltransferases identified, and similarly using SAMs as the universal methyl donor. To provide some background on these methyltransferases, the exponent attached to the m means that the methyl group is attached to the corresponding number of carbon on a cytosine or adenosine, depending on what follows. It is crucial to note that DNMTs in general are not limited to genes as they can affect other types of genetic information, such as rRNA, tRNA, mRNA, and various noncoding RNAs.

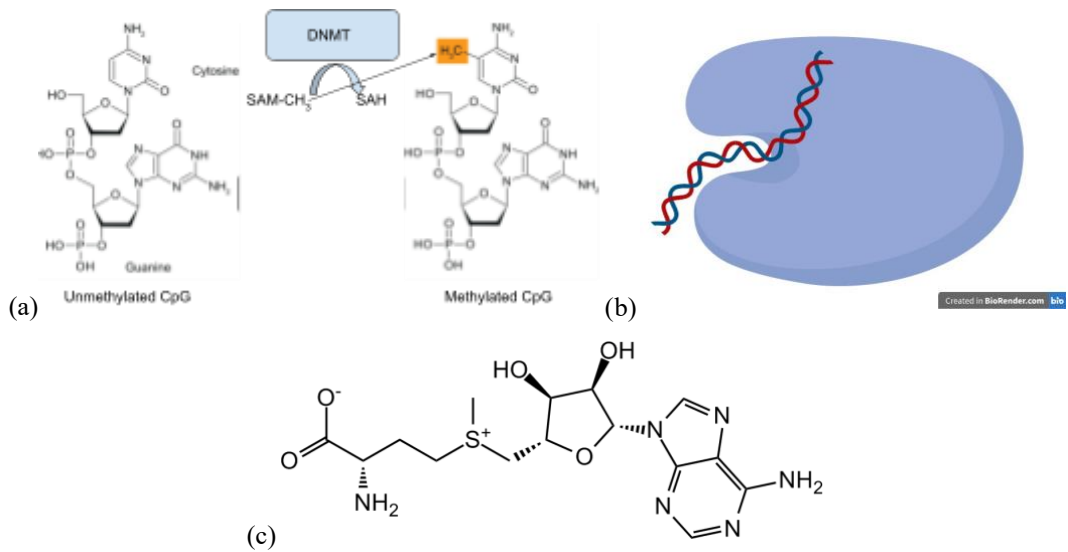

*Figure #2: (a) Starting from the left to the right, the "unmethylated CpG" contains the nucleotide cytosine and guanine, hence the CpG. This is consistent with the statement that one of the sequences that DNA methyltransferases typically act on is CG. The visual representation of the description then goes on to symbolize the transfer of a methyl group from the SAM, creating S-Adenosyl-L-homocysteine (SAH), a demethylated SAM along with other minor tweaks of its chemical structure. Finally, it is seen that the methyl group has been successfully attached to the nucleotide, which likely begins to exhibit reactions in its transcription. (b) The protein rendering of DNMT1, a common DNA methyltransferase present in many eukaryotic cells concerning epigenetic functions. The figure also shows the binding site of the DNA strand that allows for the methyltransferase to securely transmit the methyl group onto nucleotides. (c) The chemical structure of SAM ($C_{15}H_{22}N_6O_5S$) is shown, and a methyl group can be identified attached to the sulfur region. This methyl group is what the DNA methyltransferases take from SAM to make SAH and a methylated nucleotide. Created with [BioRender.com](BioRender.com)*

## *DNA Methyltransferase Inhibition*

While DNMTs are prevalent in nearly all living cells, their inhibition can still occur, which does so through mechanical means: acting on the DNMT protein itself while situating itself on nucleotide bases. In one such case, 5-azacytidine (5-Az) is an irreversible inhibitor decreasing DNMT activity, leading to a lack of methylation at certain parts of the genome, resulting in changes to cells such as suppressed tumor growth and plant phenotypic changes, including dwarfism, early flowering, and inhibition of vegetative growth.

5-Az, its brand name Vidza©, is a drug commonly used as a type of treatment for myelodysplastic syndrome, myeloid leukemia, and juvenile myelomonocytic leukemia, as well as tumor suppression. More broadly, 5-Az can selectively activate gene expression in eukaryotic cells (which is how its tumor suppression properties work) and disrupt cell differentiation under certain conditions. It works by being randomly incorporated into the DNA strand (typically the cytosine nucleotide since DNMTs are more likely to target cytosines), as roughly depicted in the diagram below, which in the presence of the nitrogen atom at the 5-position of azacytidine, results in a covalent irreversible complex for the DNMT, subsequently leaving no space for the DNMT to carry methyl groups to transfer to the nucleotides. This leads to the hypomethylation of the DNA strand, depicted as the trapped enzyme DNA adduct. (Yang et al., 2022b)

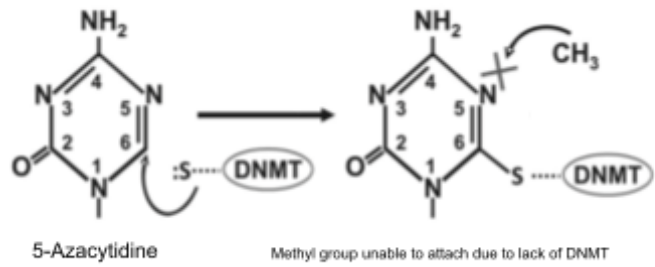

*Figure #3: DNMTi 5-Az is attached to the cytosine nucleotide of the DNA segment (not shown). It then regulates that DNMT and blocks the attachment site of cytosine such that it cannot perform its function of securely attaching methyl group.*

## *pRNAs and PAPA on Transcription*

There are complex epigenetic regulatory mechanisms that regulate rRNA genes (rDNA) through the transcription of ncRNAs, namely pRNA and PAPAS. pRNA is synthesized from the RNA polymerase I promoter in rDNA and it operates as a major silencer of rDNA loci in several ways. The first method is pRNA's recruitment of DNMT3b, which can induce de novo DNA methylation followed by heterochromatinization (Figure 2A). In addition, pRNA interacts with Histone Deacetylases to promote histone deacetylation and NoRC to enhance nucleosome sliding thereby regulating rDNA transcription (Figure 2B). Conversely, PAPAS acts as a large regulatory lncRNA produced through antisense transcription by RNAPII traversing through the gene body, promoter, and enhancer regions of rDNA. It plays an important role in growth-arrested cells by inhibiting rDNA transcription because it recruits Suv4-20h2 at its promoter which results in H4K20me3 marks deposition leading to chromatin compaction (Figure 2C). Similarly, under heat stress, PAPAS interacts with the CHD4/NuRD chromatin remodeling complex to facilitate nucleosome shifting hence helping inhibit rDNA transcription.

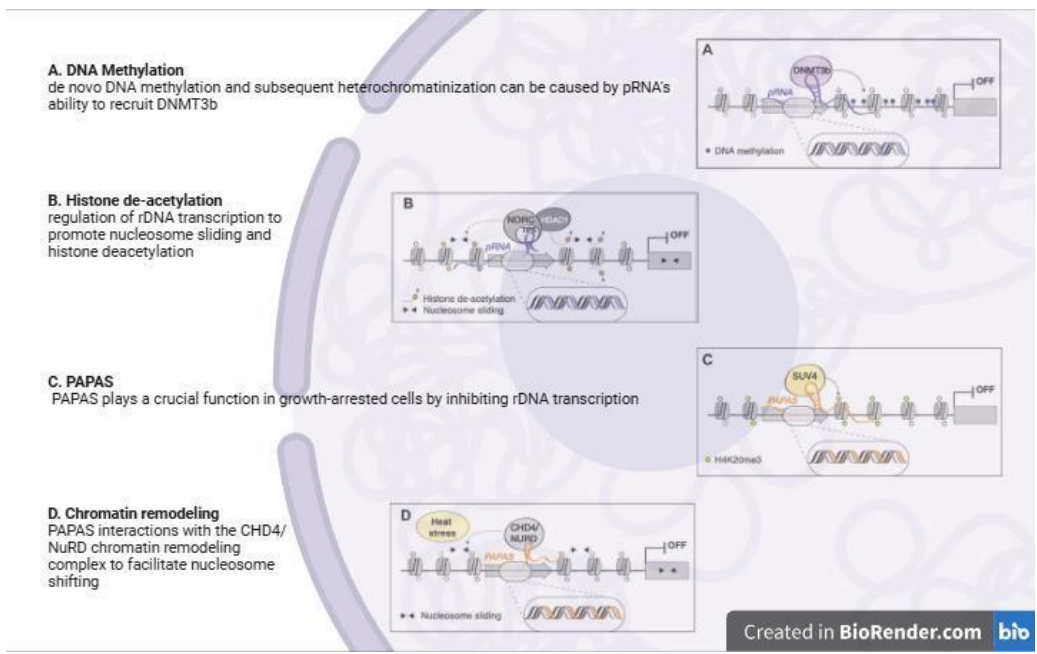

*Figure #4*: The epigenetic control of rRNA (Mangiavacchi et al., 2023c)
*Created with BioRender.com*

In conclusion, the complex means by which ncRNAs like pRNA and PAPAS influence rRNA gene expression exemplify the complex mechanisms of epigenetic regulation in biological processes. These non-coding RNAs specifically control rDNA transcription, essential for cellular homeostasis and sensitivity to environmental stimuli by coordinating chromatin modifications and nucleosome dynamics. Ongoing studies investigating these regulatory networks may lead to treatment options for diseases with dysregulated gene expression and genomic instability that have the potential to discover new therapeutic strategies targeting epigenetic vulnerabilities.

The interactions of ncRNAs extend to proteins, with one such case involving three types of proteins in charge of controlling the alteration of RNA metabolism. These three protein categories—writers, readers, and erasers—are further connected in their role in DNA methylation, particularly through their involvement with the core MTase heterodimer complexes related to m6A modification. This modification involves methyltransferases (DNMTs) such as DNMT3 and DNMT14. Specifically, DNMT3 acts as a SAM-dependent methyltransferase with catalytic properties, while DNMT14 serves as a binding platform for RNAs. (Kazimierczyk & Wrzesinski, 2021e) . The complex formed by these two methyltransferases interacts with many different factors such as Wilms' tumor 1-associating protein (WTAP) and zinc finger CCCH domain-containing protein (ZC3H13). The WTAP protein is directly correlated to m6A modification levels and a deletion of this protein will result in a decrease in m6A modification levels and as a result, cease DNMT3/DNMT14 complex activity. 6-methyl adenosine has various modification sites and they are recognized by "reader" proteins. One protein family that is commonly an m6A reader is the YT521-B homology domain-containing protein family which includes YTHDC2. This specific protein regulates the chromosome-silencing effect on lncRNA XIST by acting as a scaffold molecule. YTHDC1 is another notable protein in this family as it is responsible for gene splicing. These 3 protein groups (readers, writers, erasers) are all interconnected and play a role in regulating each other.

## Systems & Case Exploration
### *Sequencing*

As introduced earlier showing the varied effects of 5-Az, the effects of it can be detected in a direct, quantitative, and precise (single nucleotide pair precision) method by determining the methylation of cytosine in the DNA. Called bisulfite sequencing, it has been the industry standard for this purpose.

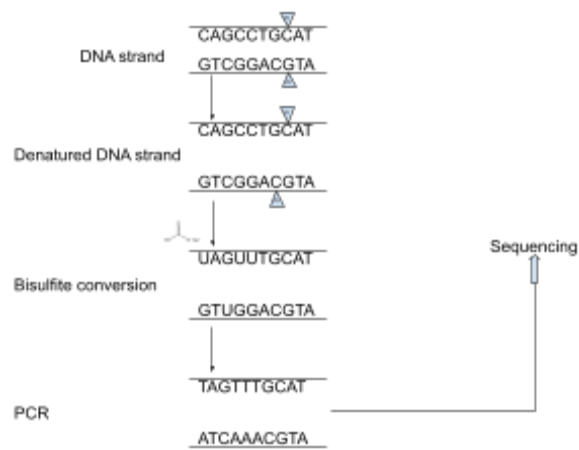

*Figure #5: A representation of the mechanism by which bisulfite conversion and its detection occur in a typical laboratory protocol.*

The mechanism of bisulfite sequencing can be summarized into 4 mechanical parts: genomic RNA preparation, bisulfite modification, bisulfite PCR amplification, direct PCR sequencing, and cloning sequencing. Cytosines in single-stranded DNA will be converted into uracil residues and recognized as thymine in subsequent PCR amplification and sequencing; however, since 5mCs are immune to this conversion, they remain as cytosines allowing 5mCs to be distinguished from unmethylated cytosines. A subsequent PCR process is necessary to determine the methylation status in the loci of interest by using specific methylation primers after the bisulfite treatment. The actual methylation status can be determined either through direct PCR product sequencing (detection of average methylation status) or subcloning sequencing (detection of single molecule distribution of methylation patterns). Moreover, bisulfite sequencing analysis can not only identify DNA methylation status along the DNA single strand but also detect the DNA methylation patterns of DNA double strands since the converted DNA strands are no longer self-complementary and the amplification products can be measured individually. It should also be noted that an RNA version of bisulfite sequencing also exists, and the main difference between the protocols is that the RNA is to be converted to a tagged cDNA, and then undergo the PCR process. Then, quantitative analysis through computer programs can be done to determine the changes of cytosine to thymine through uracil by bisulfite conversion. (Li & Tollefsbol, 2011)

Even as the current solution works for basic testing of broad averages of cytosine conversion, Dai et al. have found a new method of bisulfite sequencing called the "Ultrafast Bisulfite Sequencing" that uses highly concentrated and purified bisulfite reagents along with a high reaction temperature to accelerate the bisulfite reaction by 13 times. Benefits of this method include decreased DNA damage, lower background noise, shortened reaction times, reduced severe DNA damage, and the overestimation of 5mC level and incomplete C-to-U conversion of certain DNA sequences. (Dai et al., 2024)

Altered recipes proposed by Dai et al. include a bisulfite solution consisting of ammonium salts of bisulfite and sulfite only, instead of solutions containing sodium bisulfite, which has been shown to not primarily act for the conversion step as its compound form takes shape when sodium ions and bisulfite ions align well to carry out the conversion. Some of the numerical benefits include a 3-minute conversion period, rather than 40 minutes, which is the average for "industry standard" bisulfite conversion. Additionally, due to the short time span, the ultrafast bisulfite condition has been rated to not produce false negative results within 10 minutes (i.e. converting cytosine into compounds beyond repair or converting methylated cytosines into uracils due to excessive time). Finally, as a cumulation of the above benefits, the problems of long incubation time used in conventional BS-seq led to the degradation of over 90% of the incubated DNA through aggressive and unnecessary conversions is mitigated. Such extensive degradation is problematic, especially for certain low-complexity DNA and cell-free DNA (cfDNA). (Dai et al., 2024)

Another sequencing method was developed from various occasions of lncRNA epigenetics affecting cell differentiation and organism development leading many laboratories to screen ncRNA to determine if there is a presence of modified nucleotides. This would allow them to study cell development and the modification patterns that arise. However, epigenetic mapping is difficult due to screening methods being unable to detect modified nucleosides. One specific method utilizing two-dimensional thin-layer chromatography is the "Site-specific Cleavage and Radioisotope Labeling followed by Ligation-Assisted Extraction and

Thin-layer chromatography"(SCARLET) method which Marek Kazimierczyk and Jan Wrzesinski describe(Kazimierczyk & Wrzesinski, 2021c).  The procedure includes site-specific cleavage, radiolabeling, ligation-assisted extraction, and thin-layer chromatography. Using this method, the specific position of the m6A residue was discovered, which is significant in studying the cellular dynamics of m6A change. The SCARLET technique begins with either total RNA or a total polyA+ RNA sample. The second stage is to select a candidate site in a candidate RNA of interest. RNase H cleavage is guided by a complimentary 2′-OMe/2′-H chimeric oligonucleotide in the third stage to achieve site-specific cleavage 5′ to the candidate site. The cut location is 32P labeled, and the 32P labeled RNA fragment is splint ligated to a 116-nucleotide single-stranded DNA oligonucleotide by DNA ligase. The sample is then digested using RNase T1/A to completely digest all RNA, whereas the 32p is not digested. The labeled candidate site remains with the DNA nucleotide as DNA-32P(A/m6A)p and DNA-32P(A/m6A)Cp, which migrate on denaturing gel as 117/116 mers. The labeled band is excised from the gel, digested with nuclease P1 into 5′ phosphate-containing mononucleotides, and the m6A modification status is evaluated by TLC (Kazimierczyk & Wrzesinski, 2021b). This method was utilized to discover modified nucleotides such as m6A, m5C, and Ψ found in coding and noncoding RNAs. Despite the success of this mapping strategy, it is inefficient and has a low throughput. Many other sequencing methods are also based on short-read sequencing and recently a new detection method called nanopore sequencing was developed. These sequencing methods are currently still inefficient but they are rapidly developing and could be used to accurately and efficiently map out epigenetics and detect the presence of modified nucleotides. Multiple sequencing studies were performed showing the number of modifications of specific nucleotides in lncRNA. The results of these tests were m6A—(13357 modifications/12348 lncRNAs); m5C—(9965 modifications/1072 lncRNAs); Ψ—(162 modifications/150 lncRNAS); I—(11726 modifications/3374 lncRNAs) (Jacob et al., 2017b), (Shafik et al., 2016c), (Szcześniak & Makałowska, 2016)

     Although not fully understood, these modifications could be key to understanding the epigenetic landscape of lncRNAs and how specific proteins regulate lncRNAs which is in itself part of gene regulation and epigenetics. The regulation of lncRNAs would thus affect the distinctive role that they play in the cell, such as gene regulation and alternative splicing. Sequencing of lncRNA and specific modified nucleotides can have many useful applications, acting as a biomarker in the development of oncogenes and understanding the various diseases that are caused by these modifications.

## *Diseases*
### *Cancer*

     The basis of oncological implications in DNA methylation and especially $m^5C$ methylation is that the phenomenon occurs infrequently in normal cells. The abnormal transcriptional silencing of tumor suppressor genes by hypermethylation of $m^5Cs$ has become an attractive and selective tumor-specific therapeutic cancer target.

     5-Az not only changes S. miltiorrhiza but also cancer; along with decitabine, the two drugs have demonstrated the ability to reduce the DNA methylating activity of DNMT at low doses, showing they can induce demethylation of epigenetically silenced genes. In vitro studies on genome-wide and site-specific methylation showed that azacytidine and decitabine significantly decreased DNA methylation in lymphoid cancer cells by 60% and 40%, respectively, while for colon carcinoma cells, it achieved nearly similar results, all under certain pre-conditions of gene methylation. With a relatively low dose of 5-Az and decitabine drugs, important proteins such as TIMP3, p15, p16, CDKN1C, and RASSF1, which are all involved in essential cellular functions such as apoptosis, cell cycle, and DNA repair, can be demethylated to produce larger amounts of these crucial molecules in treating cancer along with many other diseases. (Yang et al., 2022)

     While the potential of 5-Az and decitabine as new epi-drugs for cancer-type diseases, some side effects have already been discovered. First, higher concentrations lead to strong cytotoxic effects, disrupt DNA synthesis, and cause DNA damage. Second, their demethylation function requires the S phase of the cell cycle. This phase enables these agents to selectively and effectively integrate into the DNA of rapidly dividing cancer cells, thereby minimizing hypomethylation in normal cells; however, azacytidine and decitabine do not share the same hypomethylation potency; azacytidine is only about 10% as effective as decitabine at inhibiting DNA methylation, so cautionary measures and precise calculations must be taken to make sure that such drugs do not build up to toxic levels since the S phase is only a part of the entire cell regeneration phases when taking both drugs at once. Lastly, decitabine is integrated into DNA after phosphorylation to block DNA methylation, while azacytidine is also incorporated into RNA. Around 80-90% of azacytidine is incorporated into RNA. This incorporation reduces tRNA acceptor activity, breaks down polyribosomes, and integrates into mRNA, inhibiting protein synthesis and enzyme induction. These effects can impact both cancerous and normal cells, another warning is to not allow for overdosing on the drug which could cause dysfunction in normal cells. (Yang et al., 2022) (Gnyszka et al., 2013) (Jin & Robertson, 2012)

     On the same topic of cancer, one of the clinical experiments sampled in this paper is A Phase II Study of Epigenetic

Therapy to Overcome Chemotherapy Resistance in Refractory Solid Tumors performed by The National Institute of Cancerologia. The clinical study treated the patients with a daily dose of a slow-release formulation of hydralazine tablets containing either 182 mg for rapid-acetylators or 83 mg for slow-acetylators. To briefly recap the activities of acelators, they are the opposites of methylators: while methylation makes the sequence more wounded together, acetylation allows for more open expression of the gene of interest. Another drug that was tested in the same clinical trial was the low-release tablets containing 700mg of magnesium valproate at a dose of 40mg/Kb. After the initial week of epi-drug introduction, chemotherapy will initiate the day after, with the same pre-study protocol regimen at which patients showed tumor progression. Toxicity was evaluated after each course of chemotherapy, with three formal rounds in total. The promoter of selected genes will be evaluated by MSAP-PCR in serum DNA before and after 7 days of treatment with hydralazine and valproate. (Candelaria et al., 2007)

   The results of this experiment proved that hydralazine and magnesium valproate were viable in providing a clinical benefit, where 80% of their patients observed reductions in global DNA methylation, histone deacetylase activity, and promoter demethylation. In cancer cells, the excessive acetylation of proto-once and oncogenes are major contributors to cancerous development. Therefore, with the observed reductions in DNA methylation, histone deacetylation, and promoter demethylation, the oncogenes that were previously expressed became inactive, therefore overcoming chemotherapy resistance. This is monumental to the further development of epigenetic medicines and treatments for cancer. The process by which hydralazine inhibits DNA methylation has been explained thoroughly by Wolff, et al., where he explains that hydralazine targets DNA methyltransferase, recalling from an earlier discussion, the enzyme responsible for transferring a methyl group from a SAM to cytosine or other nucleotide bases. Transitional cell carcinomas, the cancer genes studied by the institute can be targeted to be methylated, then the transitional cell cancer (TCC) genes will have lower transcription rates, treating the aggressiveness of newly developing cancerous cells. Hydralazine affects the expression of genes by regulating its ease of access to transcription factors and RNA polymerases, but since it interacts with DNA methyltransferases rather than editing the direct base pairs of the gene, it is considered an epigenetic factor. Magnesium valproate acts to regulate deacetylase histones in chromosomes, which removes acetyl groups from histone proteins, creating heterochromatin with low transcription efficiency (Wolff, et al). The last medicine, valproic acid was used to inhibit histone deacetylases, as it could be used to increase the transcription of tumor suppressor genes, which have been shut off due to cancer, creating the effects of cancer resistance. In essence, hydralazine and valproic acid have almost opposite effects and can be artificially engineered so that scientists perform hydralazine methylation onto unwanted base pairs, and use valproic acid on wanted genes. This kind of epigenetic manipulation of the activation of genes is especially significant in the realm of genetic diseases because medical professionals can direct specific genes, such as tumor suppressors and methylating disorder-causing genes. Additional applications of this may include treating patients with diseases that are caused by an overactive repressor, inducer, or any other processor of product molecules. By methylating these genes, these molecules can be reduced in function, and conversely, by acetylating a desired disease, repressing genes can help reduce the effects of the genetic disorder. Using epigenetic mechanisms to control cancer growth and chemotherapy resistance is a large step in non-invasive treatments of cancer, which previously required major surgical operations when chemoresistance is high, but with new methods to overcome chemoresistance, traditional chemotherapy associated with the new drugs will mitigate the risks from traditional cancer treatments. (Candelaria et al., 2007)

*Major Depressive Disorder*

   Traditional antidepressants are known to have resistance within the human body, as miRNAs, specifically brain-derived neurotrophic factor (BDNF) -related miR-132, influence antidepressant response heavily by disrupting antidepressant substrate binding (Tiao-Lai Huang). The Kaohsiung Chang Gung Memorial Hospital in Taiwan published the study Epigenetic Regulation of Brain-derived neurotrophic factor (BDNF) in Major Depression provides ample experiment data with the method to interview for Diagnostic and Statistical Manual of Mental Disorders (DSM-IV) criteria from 80 major depressive disorder (MDD) affected people. The data of BDNF histone modification in all subjects will be collected and the mechanism of epigenetic regulation of BDNF in major depression will be discussed. The data that they observed is listed below as a chart. (Duclot & Kabbaj, 2015)

|  | Healthy Subjects | Major Depressive Disorder Patients Before Treatment | Major Depressive Disorder Patients After Treatment |
|---|---|---|---|
| BDNF promotor I acetyl-H3 (units in relative quantification) | 3.5610 | 0.3124 | 0.4321 |
| BDNF promotor I acetyl-H4 (units in relative quantification) | 2.6346 | 0.3784 | 0.4659 |
| BDNF promoter IV acetyl-H3 (units in relative quantification) | 1.6356 | 0.4059 | 0.4303 |
| BDNF promoter IV acetyl-H4 (units in relative quantification) | 2.2795 | 0.4676 | 0.4988 |
| BDNF promotor X acetyl-H3 (units in relative quantification) | 3.8960 | 0.2564 | 0.3640 |
| BDNF promotor X acetyl-H4 (units in relative quantification) | 1.7433 | 0.2105 | 0.2540 |
| Overall BDNF Levels of MDD Patients (ng/mL) | 7.9386 | 5.6014 | 6.2803 |

*Figure #6*: Histone Modification and BDNF levels of MDD patients before and after treatment. For units in relative quantification, the higher the number, the more histone modification there was, and for BDNF levels, it is directly representative of the amount using ng/mL levels, healthy amounts of BDNF levels depicted in healthy controls.

  It is seen that there is a difference between pre-treatment and post-treatment in both BDNF promotor modification states and the overall BDNF levels in patients. Note that while in both pre and post-treatment, the numbers are all significantly apart from the healthy control, the addition of the epi-drug BDNF-related miR-132 allowed the levels of histone modifications and BDNF levels to approach the levels of that of the healthy control.

  MDD is a condition that has been historically caused by gene-environment interactions rather than being determined by genetic sequence alone, with it being linked to altered gene expression from the hypermethylation of BDNFs. The data shows that there is a definite increase in the BDNF promotors generally after the treatment of this experiment, which means that depression levels are likely reduced. Recall that BDNF is regulated by methylation by CpG segments in its sequence, cytosine and guanine with a phosphate group in between. Therefore, when BDNF promoters are active, histone modifications occur so that the CpG and BDNF sites are affected. We see that the levels of BDNF promoter I through IV grow to healthy controls after the treatment. The proof that BDNF levels are linked to depression gives way to a new type of diagnosis by biomarkers in genetic sequences. Regarding the problem of antidepressant resistance, we first need to know how classical antidepressants work, such as selective serotonin reuptake inhibitors (SSRIs), serotonin and norepinephrine reuptake inhibitors (SNRIs), tricyclic antidepressants (TCAs), inhibitors of monoamine oxidase (MAO), and lithium, upregulate, or increase the levels of BDNF in blood serum and brain, which has been linked to lower levels of depression metrics. However, these classic methods have been shown to have limitations on miRNAs, as they interfere with signaling pathways and proteins within them, which can be controlled by the attachment of miRNAs at transcriptional and translational levels. (Duclot & Kabbaj, 2015)

*Homeostasis & Metabolism*

Even the use of lncRNAs, which often have genetic causes in the human body can amount to sophisticated disease treatment patterns. Functional annotations of the top 10 ranked lncRNAs using the Database for Annotations, Visualization, and Integrated Discovery tool (DAVID) showed that each was linked to certain functional terms or descriptions in annotation. For example, out of the total 10, the lncRNA of rank 9, LOC440896 is reported to have a sequence feature of the putative uncharacterized protein FLJ45355. Thus, IGF2-AS refers to the insulin-like growth factor 2 antisense gene protein and sequence variant. DUX4L3 genes are enriched in composite; compositional bias includes Ala-rich and Arg-rich regions, and DNA binding region. The lncRNA DUX4L3 has been credited to different gene-ontology terms among them are nitrogen compound metabolic process, biosynthesis, regulation of the biological process, regulation of metabolic process, cellular metabolism, and biological regulation.

In addition, the UCSC_TFBS algorithm from the DAVID was applied to analyze protein interactions, particularly transcription factors and their target genes' sets. Among the top 10 ranked lncRNAs, four are involved in protein interactions and have functions linked to transcription factors, that include, CXCR2P1, HAS2-AS1, DUX4L3, and LOC440896. The functional annotations linked to the top 10 ranked lncRNAs are depicted in Table 3 (Sathipati et al., 2019)

Furthermore, they performed the COXPRESdb analysis to obtain co-regulation links between the genes to determine the co-expression of genes among the top-ranked lncRNAs. For example, LOC440896 is directly connected to four co-expressed genes: Included among them are cytochrome receptor-like factor 2 (CRLF2), spermatogenesis associated 24 (SPATA24), uncharacterized locus 644090, and RAN binding protein 3- like (RANBP3L). In fact, the proteins based on these genes are a part of the Jak-STAT signaling pathway and cytokine-cytokine receptor interaction. Moreover, LINC00632 is interacting with stathmin-like 4 (STMN4), myelin-associated oligodendrocyte basic (MOBP), and kinesin family member 1A (KIF1A) Gene, and IGF2-AS is connected to like-glycosyltransferase (LARGE), nyctalopia (NYX), and the D site of albumin promoter (albumin The gene co-expression networks for these lncRNAs illustrate the probabilities of them having regulatory roles and connections, which are possibly significant for the NB pathogenesis and searching for new therapeutic targets.

Thus, to determine the levels of these three lncRNAs in NB patients, bioinformatics and subsequent validation by lab analysis of 88 human NB samples were conducted. The findings of this study assist in furthering the knowledge of the functional competencies and relations of these lncRNAs to justify them as biomarkers and treatment options for Neuroblastoma.

*Chronic Pain*

One of the biggest challenges for many people on the planet is dealing with pain, which creates a $560 to $635 billion money drain every year in the United States. (Gaskin & Richard, 2012) Acute pain, pains that last the duration of the invigilator, is essential to protecting the human from danger. However, when chronic pain develops, not only is it, to put it mildly, a nuisance, but it also serves no purpose in protecting the human from harm. ncRNAs offer a potential solution to this problem, which acts by disrupting the signal pathways and production in neurons, which will be elaborated on later. It must be noted that the development from acute to chronic pain is incredibly complex, including amplified pain response due to regular stimulus that occurs in peripheral neural encodings of impending or actual tissue damage. Chronic pain also changes neurons and glial cells in the central nervous system (CNS). Mauceri D. states that "typical symptoms may be allodynia pain from a stimulus that usually does not invoke pain signals, defined as pain perception in response to innocuous stimuli, and hyperalgesia excessive pain from a normal stimulus, in which painful stimuli are perceived at greater intensity." In neurons, which are mature and differentiated cells that can not undergo mitosis or postmitotic cells, epigenetic erasers for DNA demethylation in neurons can be actively controlled (having specific molecules to demethylate) by Growth Arrest and DNA Damage-inducible (GADD) 45 proteins, and ten-eleven translocation (TET) family of dioxygenases. This discovery is important in terms of chronic pain because there is a new possibility to alter the gene expression and function of post-mitotic adult neurons, defining the new term: neuroepigenetics. GADD 45 proteins are proteins that up-regulate (increasing the response to a stimulus) DNA damage and other problems to growth arrest and apoptosis. GADD 45 proteins are very crucial in controlling chronic pain, as they provide positive feedback to pain, which would likely develop into hyperalgesia. In most cases, the methylation of a gene will cause GADD 45 protein to be expressed in lesser numbers. (Mauceri, 2022)

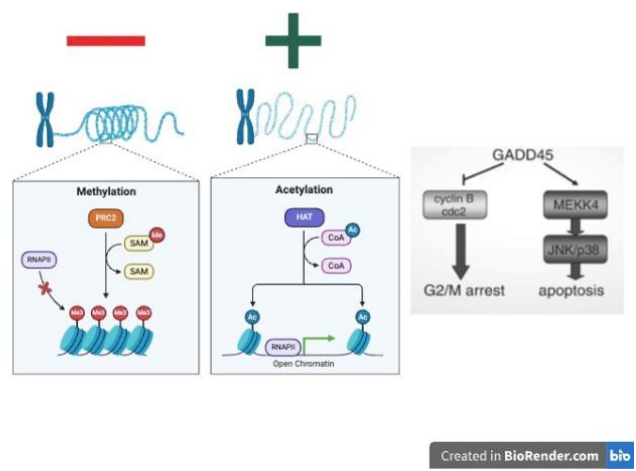

*Figure #7* Difference in the function of methylation and acetylation in the expression of GADD 45 proteins. The typical result of the epigenetic change is observed where methylation causes the restriction of transcription while acetylation opens up the DNA for transcription.

Another important regulator involved in chronic pain, TETs, however, are active contributors to demethylation. Activated by inflammation, TETs can take off methyl groups of genes. These enzymes can also be controlled by various methods of methylation like the addition or removal of DNA methyltransferases. In essence, GADD and other DNA methyltransferases and TET enzymes work in opposite directions to control the up-regulation of DNA damage to growth arrest and apoptosis. miRNAs, under the group of ncRNAs, affect neuropathological diseases by arresting the further translations of mRNAs in the central nervous system, as well as helping neurons wire into the established neuron systems and synapses. These neurons in the central nervous system play the most crucial role in pain signaling. To trace the pathway of a signal through the central nervous system, we first start in the soma region of the neuron with the nucleus, of the dorsal root ganglia (DRG) and surrounding structures, a region of the brain used to transmit peripheral nerve impulses to the CNS, containing the dorsal horn of the spinal cord. This region of the brain is susceptible to hyperalgesia and its development into chronic pain. Imagine a telephone game, where each person hears what the previous person said and is tasked to pass it on. It is especially prone to mistakes as you can only whisper in their ears, only the mistakes cause massive amounts of pain. miRNAs can be engineered to have a large impact on this transfer. miRNA in the CNS has been shown to have mixed results. Some say that miRNAs up-regulate, have no effect, or even down-regulate pain reception. miRNAs target specific genes associated with gene sensitization, such as those utilizing calcium, potassium, and sodium channels. Additionally, miRNAs can influence key signaling elements and pain mediators, including numerous molecules like the brain growth factor. lncRNA has also been shown to have some effects on chronic pain. Mauceri (2022) states that "…first identification of the voltage-gated potassium channel (Kv) Kcna2-AS-RNA, a lncRNA defined as an endogenous antisense transcript targeting the Kcna2 mRNA in neuropathic pain…LncRNAs have a different mode of action, as they might interact and interfere with miRNA and their processing or directly target specific molecules involved in pain processing…". By using ncRNAs to regulate and repair DNA and GADD45 proteins to upregulate pain response, and TETs to demethylate pain molecules, patients with chronic pain can reduce, or even prevent it without taking traditional painkillers or other drugs that have many side effects such as addiction and related irritation or drowsiness. (Mauceri, 2022)

*Agricultural Applications*

Knowing the mechanisms and effects of DNMTs, SAMs, and DNMTis, a study explores their application and seeks confirmation of those item's effectiveness. This study allows readers to see computational proof for the effects of such parts. When environmental stressors such as high salinity are placed upon wheat cultivars, they have different gene expression patterns, which were determined to be caused by the methylation in the promoter region. This example suggests that stress can cause greater salt tolerance through the increased expression of stress-resistance genes by epigenetic means. The researchers in this study investigated the metabolite production by *Salvia miltiorrhiza* (red sage) by changing the amount of 5-Az exposed to the plant, which ultimately changed the expression of hairy roots on the plant. The researchers used the plant's secondary metabolites (such as the liposoluble dihydrotanshinone I, Tanshinone I, Cryptotanshinone, Tanshinone IIA, and Tanshirone IIB, and water-soluble Salvianolic acid B and Rosmarinic acid, which are water-soluble) levels and reactions to quantify plant stress resistance. (Yang et al., 2022)

In the study, the researchers added a solution of 5-Az with varying molarity to the red sage to observe the drug's effects. After 28 days of treatment, the fresh and dry weights revealed significant growth retardation in the hairy roots of the red sage plant exposed to 5-Az. The fresh and dry weights of the roots were reduced by about 40% with 12.5 and 25 μM 5-Az in the medium. The root mass decreased by more than half when the medium contained over 50 μM 5-Az. Additionally, the researchers identified a noticeable red color change was observed in the roots treated with 5-Az, which may have been caused by the accumulation of tanshinone components. The following flow chart summarizes the mechanistic results of the study:

Addition of 5-azacytidine → lower levels of DNA methylation in sequences related to root hair weight → the sequence is allowed to be expressed → increased protein production (hence red tint) → produced proteins retard the growth of the hairy root

The quantitative analysis of this result is recorded by measuring the amount of secondary metabolites, which are active and critical components of the plant's hairy root structure and function. It was concluded that the addition of 5-Az had affected the levels of those critical components of the plant's hairy root-building blocks. A trend was noticed, where at concentrations of 5-azacytidine over 50 μM, the amounts of the liposoluble critical components of hairy roots nearly doubled or grew fivefold at 50 μM and 75 μM respectively, leading to the retardation of root hair growth. It was also noted that for water-soluble components of the plant's hairy roots, the amount generally decreased with some insignificant differences throughout. A chart detailing the specific amounts is provided in Fig. #. (Yang et al., 2022)

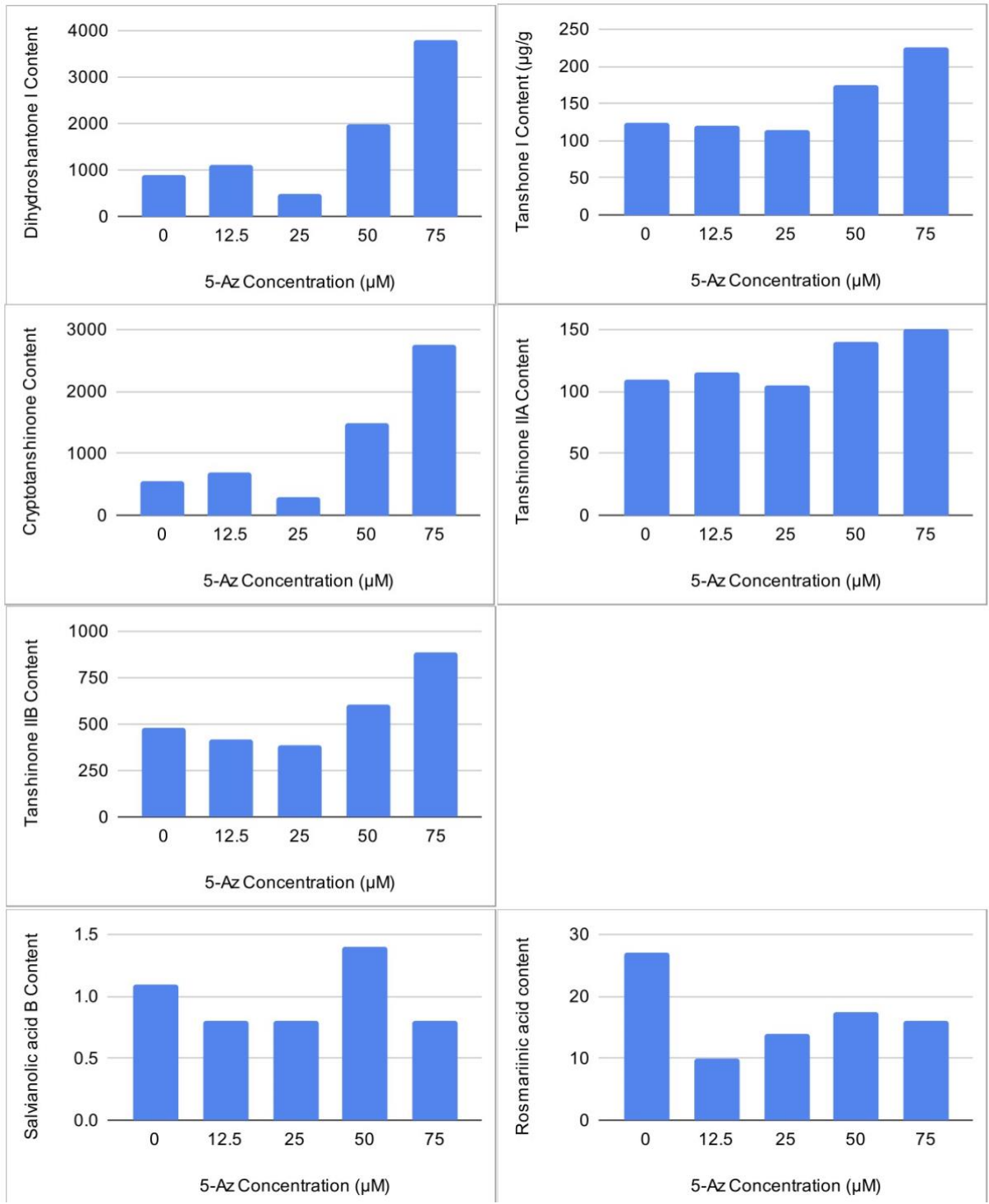

*Figure #8: Amount of several secondary metabolites against DNMTi 5-Az concentration. It is evident that in tanshinones, the amount increases with the increase in 5-Az while the amount of salvianolic acid and rosmarinic acid are more ambiguous in the trend of their changes in the context of increasing 5-Az.*

     A more definitive reading of methylation activity by the 5-Az was conducted by the same researchers throughout the entire genome of the hairy root cell. The researchers studied a 75 μM 5-Az solution where the hairy root cells were placed. Through the use of bisulfite conversion and next-generation sequencing, the methylation level of cytosine on the copalyl diphosphate synthase (CPS), an active enzyme of gibberellin biosynthesis, promoter was altered with the 5-Az treatment, with 145 cytosines detected, 51 cytosines lost their methylation (which were all transcription factor binding sites), leading to a near 65% decrease in methylation of the CPS promoter region. This significant decrease in the methylation of the promoter region increased the production amount of each of the transcription factors associated with the demethylation. While gibberellins do

cause root growth, their main targets are to grow stem cells, rather than root cells; with the increase of other secondary metabolites produced by the decrease in methylation, a net negative growth was still observed.

Other non-drug epigenetic factors are present as well. It is a well-established fact that heavy metals retards plant height [(Kiran et al., 2022)](#) through methods such as but not limited to an inhibition in root growth that disturbs the homeostasis of the plant by disrupting the root growth and other water and nutrient absorption, which results in a lowered productivity of biomass production. However, another mechanism at play is not typically thought of, which is that of an epigenetic basis.

Heavy metal agents such as cadmium, lead, mercury, aluminum, and arsenic have been shown to cause various episodes of epigenetic interactions in plant species. These metals often target global and site-specific DNA methylation, histone tail modifications, and epitranscriptome modifications that result in devastation to the plant, including a puncture to the mineral homeostasis, which could then cause other ailments such as decreased photosynthetic productivity, cell division problems, unnecessary stresses, genotoxic effects, which all consequent cumulatively to the retardation of a plant, if not the death of the plant under extreme conditions.

On a smaller scale, plants also undergo the typical epigenetic changes observed in other organisms, where their DNA cytosine nucleotides find themselves, very commonly in plants, with methylation of the sequence in the promoter region of the genome. Apart from the typical processes that plants have in common with nearly all prokaryotic and eukaryotic organisms, plants (and some other eukaryotic organisms) can also be epigenetically changed through epitranscriptomic means. The mechanism behind epitranscriptome generally resembles that of epigenetic mechanism, except for the fact that the changes to the DNA backbone are instead done to the RNA backbone, typically occurring on cytosine and adenine sequences of the RNA.

Next, a more detailed exploration of the effects of specific heavy metals on specific plants is provided as a type of case study on these phenomena.

*Cadmium Treatment*

Cadmium has been regarded as an agent for headaches and flu-like symptoms, swelling of the throat, and tingling hands in humans, and plants, significantly reducing their growth by damaging membranes through lipid peroxidation and causing disturbances in chloroplast metabolism by inhibiting chlorophyll biosynthesis and reducing the activity of enzymes involved in CO2 fixation, where such effects act on the proteomic level causing the devastation to the plants. New research, however, has identified new ways that cadmium interacts with the epigenomic level of molecular biology, and such discoveries will be covered briefly in the following organizer. [(Benavides et al., 2005)](#) MSAP-PCR refers to a method of polymerase chain reaction that is methylation-sensitive amplification polymorphism, which essentially gives a brief and overall view of the ubiquity of methylations of the DNA of interest. [(Kiran, 2022)](#) [(Chmielowska-Bąk, 2023)](#)

| Thorn Apple | MSAP-PCR analysis showed that both soil and foliar applications of Cd result in increased DNA methylation. |
|---|---|
| Isoetes sinensis | Cd and Pb triggered hemi-methylation, which is cytosine methylation on one DNA strand. Conversely, a reduction in full methylation, involving both DNA strands, was seen in plants exposed to these metals. |

| Soybean | Cd did not influence the total DNA methylation level, as measured by ELISA assay (which assesses the abnormal expression of proteins or other organic products), either immediately after treatment or after seven days of recovery. |
| | In soybean cell suspensions, cadmium triggered the accumulation of H2A histone protein. The epigenetic changes were apparent in the mRNA produced by the cell, with |
| | barley showing increased mRNA levels after a 7-day cadmium exposure, which generally caused adenosine hypermethylation. Out of 435 differentially expressed genes, 319 genes exhibited increased methylation and RNA abundance. |
| Kenaf | The hybrid F1 showed greater Cd tolerance and the lowest DNA methylation compared to the parent cultivars Fuhong 992 (CP085) and P3A (CP089). Among genes with differential methylation, those involved in Cd response like leucine-rich repeat receptor-like kinases (LRR-RLK), a protein of the NRT1/PTR family (NPF2.7), and DEAD-box ATP-dependent RNA helicase 51 (DHX51) were hypomethylated. Conversely, genes encoding trehalose-phosphate phosphatase (TTP-D), NADP-dependent malic enzyme (NADP-ME), and NAC-domain containing protein (NAC71) exhibited hypermethylation. These genes also showed varied expression in response to Cd, without a clear connection to their methylation patterns. LRR-RLK and NAC71 were down-regulated while NPF2.7, DHX51, TTP-D, and NADP-ME were up-regulated. |
| *Noccaea caerulescens* | Another study assessed the responses of the metal hyperaccumulator *Noccaea caerulescens* and the non-hyperaccumulator *Arabidopsis thaliana*. Although *N. caerulescens* absorbed more than three times the amount of Cd compared to *A. thaliana*, the latter exhibited significantly more DNA damage and higher levels of the DNA oxidation marker 8-oxo-dG. Methyl-sens comet assay results showed that Cd-induced DNA methylation in *N. caerulescens*, accompanied by increased MET1 expression. |
| *Arabidopsis thaliana* | MSRP-PCR results indicated that Cd treatment increases the methylation of external cytosines, with no significant change in internal cytosines. In contrast, *A. thaliana* showed decreased DNA methylation, lower MET1 expression, and induction of DRM2. The increased DNA methylation observed in the hyperaccumulator might contribute to DNA protection and higher metal tolerance. |

*Figure #9*: A chart presenting the extent of methylation for epigenetic-related compounds' levels in the corresponding plants. Effects were studied primarily through cadmium, and its effects on average are found to cause methylation, thus the inhibition of certain transcriptomic molecules' production.

*Arsenic Treatment*

Arsenic exposure leads to epigenetic changes, such as reduced DNA methylation in the old fronds of the Cretan brake fern (Pteris cretica L.). However, young fronds showed no change in methylation regardless of arsenic levels. (Chmielowska-Bąk, 2023)

*Rice in Various Epigenetic Factors*

Whole genome bisulfite sequencing (WGBS) of rice (Oryza sativa L.) identified almost 2,400 differentially methylated regions (DMRs), with hypermethylation particularly evident in the upstream regions of genes. Functional analysis the researchers performed indicated that the methylated genes were primarily involved in "cellular processes," "metabolic processes," and "response to stimuli."

For instance, in *Arabidopsis* species, exposure to Cu and Ni led to increased tolerance in progeny to these metals and also to methyl methanesulfonate and salt stress (Rahavi et al., 2011). In rice, heritable changes in methylation patterns were noted. Plants treated with Cu, Cd, Cr, and Hg exhibited DNA hypomethylation in the CGH context linked to TE and protein-coding genes. This DNA methylation modulation was accompanied by altered expression of genes related to chromatin regulation — primarily reduced expression of methylation-related genes (MET1-2, CMT3-2) and increased expression of demethylation-associated genes.

To focus on a specific plant, the Oryza sativa ssp. Japonica was treated with 5-aza-deoxycytidine (decitabine), a known epigenetic inhibitor, and grown in the field for over ten years. The researchers of this study aimed to create an environment in which epigenetic qualities can be inherited and passed on to the generations afterward. The researchers finding significant proof of the possibility of inheritable inheritance would bring many surprises to high school students, or even college students, who are told that Larmack lost to Darwin in determining the way that evolution happens. This study gives Larmak some credit, where he reached the right conclusion through the wrong methods. (Akimoto et al., 2007)

The method of the study was to treat the rice seedlings with 5-aza deoxycytidine (azadC) and through the course of the study, which was >10 years, they observed the effects of that drug on the growth of the crops. After the drug was administered and the growth period of the generations had elapsed, many of the plants had died due to a myriad of reasons, with some being linked to the addition of the drug, but not being of critical importance in their study. Of the remaining crops, one healthy subgroup of the batch had an average of a 28% decrease in stem length and an even more surprising result, that the heading of the rice plant became decoupled with the time that plant development occurs. Typically, plant development in the rice plant's stems region stops after the heading of rice, which is the time when the plants often develop seeds and flowers, and those flowers "head" up, but with the addition of decitabine in the previous generation, the observed generation's heading time showed the anomaly. (Akimoto et al., 2007)

By treating the cells with decitabine, an artificial methylation erasure resulting in significant demethylation could have been detrimental to survival because the majority of treated seedlings were fatal, and only those that were affected by non-essential genes might have survived. Dwarfism and illness obstruction are such models, recommending that demethylation has happened at restricted genomic districts in these lines. This mechanism is one part of the puzzle of understanding epigenetic inheritance. Whether such specific demethylation at a specific quality locus happens under normal circumstances is exceptionally compelling for epigenetic guidelines.

It was in this way proposed that methylation/demethylation happens under regular circumstances and that heritable epigenetic changes assume a critical part in development. This idea is supported by the present finding, which demonstrates that methylation can flexibly adjust gene expression, allowing plants to acquire or lose heritable traits provided that the methylation patterns of the corresponding genes are maintained. (Akimoto et al., 2007)

The key findings of the treatment were quite unexpected to the academic world as well as the researchers, who found that inheritance was possible through environmental means. This is on the side of the idea of Lamarckian legacy, it is heritable to propose that procured attributes.

*Brassica rapa*

In Brassica rapa plants, small RNAs (smRNAs) are crucial epigenetic factors influencing the plant's epigenetics, impacting gene expression both at their origin and in non-cell-autonomous gene regulation.

The analysis revealed that although significant changes in gene expression were observed in the embryo and endosperm tissues, few changes were transferred to the leaf tissue of the progeny at the 2-week post-germination stage (116 differentially expressed genes, q < 0.05, Figure 1). This indicates that gene expression changes are reversed either during the final stages of seed maturation or later during seed germination and plantlet development. (Bilichek, 2015)

| miRNA | Gene ID of putative target | miRNA's mode of action |
|---|---|---|
| bra-miR167 | Bra015704 | Translation inhibition |
| bra-miR167 | Bra002277 | mRNA cleavage |
| bra-miR167 | Bra025064 | mRNA cleavage |
| bra-miR167 | Bra005019 | Translation inhibition |
| miR168 | Bra032254 | mRNA cleavage |
| bra-miR171a-1 | Bra039431 | Translation inhibition |

*Figure #10: The effects of certain miRNA on certain genes of the Brassica rapa plant, which overall has effects on the transcriptomic and translatomic levels of gene regulation of eukaryotic systems.*

## DISCUSSION

Epigenetics holds immense potential for advancing our understanding and treatment of a wide range of diseases. Finding the nexus between epigenetics and ncRNA will finally allow us to connect the two and deepen our understanding of their complex interplay. The reversible nature of epigenetic modifications renders them attractive intervention targets in a therapeutic context. For example, epigenetic therapies including DNA methylation inhibitors and histone deacetylase (HDAC) inhibitors are currently in use or being studied as a treatment strategy for cancer to do just that; they reactivate silenced tumor suppressor genes while shutting off the activity of oncogenes, which Gnyszka (2013) supports although casting suspicion on the long term effectiveness of drugs such as decabitine and calling for the need of new and more direct targeting drugs. In addition to the field of oncology, epigenetics is anticipated to have significant implications in neurodegenerative diseases as this branch makes it possible for one to alter abnormal marks and thus normalize gene function which could significantly reduce disease progression. More traditionally, chemotherapy drugs and treatments become much more effective in certain individuals with high levels of chemoresistance through epigenetic therapy of the patient which was studied by Candelaria et al. (2007). Moreover, epigenetic markers can be used as diagnostic or prognostic biomarkers for the identification of potential earlier treatment onset and precision medicine. Moreover, the emerging technologies to edit epigenetic marks (e.g., CRISPR-based epigenome editing) equip us with an additional enrichment that we could co-integrate in our armamentarium to introduce spatially targeted changes in gene expression without perturbing the DNA sequence. The more we learn about the epigenome, the closer we might get to having this science as a game changer in the medical field.

Epigenetics and noncoding RNA interact in a highly complex manner to form a regulatory network fine-tuning gene expression and thereby modulating cellular homeostasis and the establishment of pathologies. In Statello et al. (2020) we discussed ncRNAs, particularly lncRNAs, are significant players in the epigenetic regulation of gene expression. They affect chromatin state and gene activity in various ways. For instance, the lncRNA Fendrr interacts with Trithorax Group (TrxG) and Polycomb Repressive Complex 2 (PRC2) proteins, influencing the deposition of active H3K4me3 and repressive H3K27me3 marks, respectively. This dual regulatory ability highlights the complexity of lncRNAs in maintaining the balance of gene expression necessary for healthy development and cellular differentiation. Fendrr's ability to modulate TrxG and PRC2 ensures that genes are either appropriately active or repressed, which is crucial for processes like the growth of the caudal lateral plate mesoderm and the organism's overall body plan. Another study where specific ncRNA signatures are determined for disease outcomes, exemplified by a neuroblastoma lncRNA signature as a prognostic factor, underscoring the potential of ncRNAs to act as biomarkers in a clinical setting.

    Tsai et al. (2022) establish that TERRA decreases strongly impair the G4 structures at TSS, such genes as MYC and VEGFA associated with angiogenesis and cell division. This decrease is associated with changes in the cellular levels of RNA species, which suggests that TERRA regulates the genes' activity by stabilizing G4s. Thus, protecting these structures allows the organization of chromatin required for appropriate gene regulation, with TERRA's assistance. This mechanism is in concordance with other talks regarding how lncRNAs such as Fendrr and MEG3 bind to chromatin-modifying complexes to control gene expression. For instance, Fendrr targets H3K4me3 and H3K27me3 marks by modulating the function of Trithorax-Group (TrxG) and Polycomb Repressive Complex 2 (PRC2) proteins. ATRX is the chromatin remodeling protein involved in resolving G4 structures and TERRA is also known to affect ATRX. Reduced TERRA level affects the binding of ATRX at G4 sites thus destabilizing the TSS of genes that are actively involved in cellular functions. This disruption can lead to the alteration in the normal functioning of tumor suppressor genes that might in one way or the other cause carcinogenesis. This aspect of TERRA's function is as follows the roles of other lncRNAs in chromatin interaction and gene regulation. For example, MEG3 promotes PRC2 binding to the chromatin site to increase the repression of genes. As in the case of the relationship between TERRA and ATRX, it aligns the role of lncRNA in sustaining genome stability and precise gene expression.

    In the context of neuroblastoma discussed by Sathipati et al. (2019a), lncRNAs act as important biomarkers for diagnosis and prediction of the progression. Among 783 investigated lncRNAs, 35 have been revealed to have a strong connection with NB patients' overall survival. These lncRNAs are oncogenetic and their increased expression correlates with the disease progression and worse survival. Thus, this study has established the role of lncRNAs in the progression of cancer as well as lending credence to the use of lncRNAs in the creation of treatments. Compiling this information with the prevailing discourse on epigenetic regulation as well as ncRNAs, this situates lncRNAs as highly versatile molecules involved in gene regulation and disease development. Similar to other lncRNAs such as Fendrr, MEG3, Xist, and TERRA, the lncRNAs identified in NB are significant to the understanding of NB cellular process and cancer gene regulation and also to the development of diagnostic markers and therapeutic approaches. The part played by lncRNAs in chromatin interaction is mentioned in both discussions repeatedly. For example, lncRNA MEG3 helps recruit PRC2 to the genomic regions to promote the process of gene silencing. Likewise in NB, LO440896, lncRNA participated with Co-Expression Network data which gives implications to the regulation of gene expression. These findings associated with chromatin interaction mechanisms and lncRNAs' roles in NB also support the roles of these molecules in epigenetic regulation. In the study on NB, DAVID or Database for Annotations, Visualization, and Integrated Discovery was used to conduct functional annotation of top-ranking lncRNAs. Most of these lncRNAs are associated with protein interactions, particularly with transcription factors that are very essential in regulating gene expression. For instance, LOC440896 physically associates with co-expressed proteins in the Jak-STAT signaling molecular pathway and cytokine-cytokine receptor binding. This functional annotation is similar to the chromatin and gene regulation aspect of lncRNAs as explained above. There are major therapeutic implications for the results.

    Kazimierczyk (2021) discusses the presence and significance of altered nucleotides in lncRNA, focusing on common

modifications such as 6-methyladenosine (m6A), 5-methylcytidine (m5C), pseudouridine (Ψ), and inosine (I). These modifications are introduced and regulated by three types of proteins: "writers" (e.g., methyltransferases like DNMT3 and DNMT14), "readers" (e.g., YTH domain-containing proteins), and "erasers" (e.g., demethylases).

The roles of the proteins described above include the regulation of the metabolism of RNA and its importance in the functioning of lncRNAs. For instance, the methyltransferase complex composed of DNMT3 and DNMT14 promotes the generation of m6A modifications. This complex regulates certain factors including Wilms' tumor 1-associating protein (WTAP) and zinc finger CCCH domain-containing protein (ZC3H13) that exert a control on the m6A modification levels and the DNMT3/DNMT14 complex. These m6A modifications are read by "reader" proteins such as YTHDC1 and YTHDC2; the former is involved in mRNA splicing while the latter is involved in the silencing of chromosomes. This mechanism is commensurate with other talks regarding how TERRA and other lncRNAs engage chromatin-modifying complexes in the processes of gene regulation. For example, TERRA stabilizes G-quadruplex (G4) and regulates ATRX binding involved in genome integrity and gene expression. Identification of lncRNAs is highly complex when it comes to epigenetic mapping because detecting modified nucleosides is very difficult. Kazimierczyk and Wrzesinski also presented an outstanding site-specific technique named SCARLET that can identify the modification, for instance, m6A. This method involves processes like cleavage in the site, radio labeling, and TL chromatography; thus, making it possible to analyze the cellular behaviors of RNA modifications. Additionally, the functions of lncRNA in NB also belong to the regulatory mechanisms described by Kazimierczyk and Wrzesinski concerning the altered nucleotides' role in lncRNA. As exemplified here, lncRNAs play more than one part in controlling gene expression for disease progression and presenting fresh antithetical approaches to treatment.

Learning more about the functions of lncRNAs and their changes in epigenetics will help to create new methods for diagnostics and treatment of diseases. Therefore, understanding specific lncRNAs that relate to the survival rates in NB would help enhance the patient's status in NB. In the same way, it is an opportunity to develop new drugs for diseases associated with changes in the epigenotype that might occur due to specific modifications to lncRNAs. The advent of better techniques for sequencing including nanopore sequencing might enhance the high-throughput efficiency of mapping epigenetic changes in lncRNAs. Several works have demonstrated changes in lncRNAs across sequencing studies, and it is suggested that lncRNA might have an epigenetic function in the understanding of the gene regulation network.

      miRNAs, under the group of ncRNAs, affect neuropathological diseases by arresting the further translations of mRNAs in the central nervous system, as well as helping neurons wire into the established neuron systems and synapses, which help to reduce the sensitivity and activity of pain sensors that are related to sources of chronic pain disorders. These ncRNAs form the foundation for future clinical trials that test for the phenotypic outcomes of such molecular behavior, which could potentially lead to alternative forms of treatments that are minimally invasive, inexpensive, and effective. Furthermore, clinical trial data by Fackler et al. (2011) suggest the targeting of epigenetic mechanisms for cancer therapy and evidence of the efficacy of epigenetic drugs in clinical settings, as noted on ClinicalTrials.gov. This view from practice is further supported by studies like Jarroux et al. (2022), who discuss therapeutic potential regarding the role of ncRNAs in regulating gene expression and their involvement in the pathogenesis of diseases published in the International Journal of Molecular Sciences.

Furthermore, Valadkhan et al. (2009) detail RNA-mediated epigenetic regulation with complex feedback mechanisms between ncRNAs and epigenetic modifications that function specifically and complicatedly in their regulatory networks.

      The molecular mechanisms of polymerase II regulation are fundamental to comprehending ncRNA-mediated gene expression regulation. These interactions are one of the major components of epigenetic regulation because ncRNAs may affect the gene function even if they do not change the DNA sequence. Another paper by Espinoza et al. attempts to unmask the complicated dynamics of ncRNA-Pol II associations, emphasizing B2 RNA. This research uses double-strand DNA conjugated with a fluorescent marker for the AdMLP area to analyze the formation of complexes and the formation of elongation complexes by native polyacrylamide gel electrophoresis. Primary works utilizing in-vitro transcription together with RNA structure

elucidation methods including RNase probing, in-line probing, and RNase footprinting were used to establish the structural and functional details of B2 RNA. The work shown in the study also depicts how B2 RNA, a component of human chromosome 21, binds to the transcription bubble of Pol II, and thus it halts the process of generating mRNA from DNA. This interaction depicts a view where the ncRNA can influence gene expression and this is an epigenetic regulation. B2 RNA is typical of the transcriptional repressor, which is the exact evidence of the involvement of ncRNAs in the regulation of the gene. This is also reflected in how the research shows that B2 RNA is regulated depending on certain conditions. The regulation of B2 RNA binding and inhibition is affected by transcription factors and chromatin states; the ChIP assays indicate this. For example, B2 RNA interacts with genes that are perceived to have restrictive histone modifications such as H3K27me3, which puts it on the map of the cell's epigenetic program. The established study reveals the following aspects of the protective interactions between pol II and B2 RNA. When generating RNase footprinting data, Pol II prevents certain nucleotides in the B2 RNA from being cleaved by RNase, especially between nucleotides 73 and 131. These protective interactions are overlaid onto the secondary structural model of B2 RNA and the exact positions are indicated on the model. The findings of this study are consistent with other discussions on ncRNAs and their involvement in epigenetic regulation. For example, previously mentioned by Marek Kazimierczyk (2021), the regulation of lncRNAs includes the relationships with writer, reader, and eraser proteins that modify nucleotides in lncRNAs. Likewise, the case of TERRA and G-quadruplex (G4) structures explains how ncRNAs are involved with the maintenance of genome and gene expression.

      The investigation of polymerase II and B2 RNA also provides a further layer of the concept to be added to the ncRNA function, where it is evident that ncRNAs are capable of acting as transcribable repressors by directly associating with the transcription machinery. This is in concordance with the overall theme of ncRNAs as multifunctional molecules that regulate genes via epigenetic processes. Studying the relations between Pol II and ncRNAs such as B2 RNA has important consequences in pathology and the therapy field. For instance, in diseases like cancer, where the gene regulation is altered, it is quite possible that by focusing on targeted ncRNA-Pol II the researchers may offer new approaches to management. The role of B2 RNA can also be context dependent which may imply that its function may be employed for personalized medicine which is treatment regimens based on epigenetic characteristics of an individual. Due to the ability to make reversible modifications, epigenetic modifications are considered ideal to form the basis for therapeutic interferences. In Mangiavacchi (2023), we examined how ncRNAs, primarily lncRNAs, are important in the epigenetic regulation of gene expression. They have multiple ways of affecting chromatin state and gene activities. For instance, Statello et al. (2020) Fendrr lncRNA interacts with Trithorax Group (TrxG) and Polycomb Repressive Complex 2 (PRC2) proteins leading to active H3K4me3 and repressive H3K27me3 marks being deposited respectively. This versatility highlights how difficult it is for lncRNAs to strike a balance between the upregulation needed for normal development and differentiation in cells. The ability of Fendrr to modulate TrxG and PRC2 ensures that appropriate genes are either on or off, which is crucial for processes such as caudal lateral plate mesoderm growth as well as overall body organization. Espinoza et al.'s study on Pol II and B2 RNA interactions provides a mechanistic example of how ncRNAs can regulate gene expression through epigenetic mechanisms. These results align with other studies that show the roles played by ncRNAs in regulating gene transcription. For example, review 1 Yenching Lin et al.'s (2022) study determines specific signatures of ncRNA associated with disease outcomes, one such example being neuroblastoma lncRNA signature as a prognostic factor, underscoring the potential of ncRNAs to act as biomarkers in a clinical setting. In the context of neuroblastoma, lncRNAs can be an important marker for diagnosis and prognosis of progression. Of these 783 lncRNAs investigated, 35 showed a strong correlation with overall survival in NB patients. All of these lncRNAs were found to be oncogenic, and high expression was associated with worse survival and advanced disease. The study, therefore, defines the role that lncRNAs play in cancer progression and thus lends support to the development of treatments using lncRNAs. Compiling information about them places lncRNAs as highly versatile molecules implicated in gene regulation and the development of diseases. Similar to other lncRNAs, such as Fendrr, MEG3, Xist, and TERRA, it is meaningful to find and identify lncRNAs involved in NB to know more cellular

processes in NB and the regulation of cancer genes and to find diagnostic markers and therapeutic approaches. The functional annotation for top-ranking lncRNAs in the study on NB was done using DAVID. Most of the lncRNAs are associated with protein interactions, especially with those transcription factors that are essential to gene regulation. For example, LOC440896 physically associates with co-expressed proteins in the Jak-STAT signaling molecular pathway and cytokine-cytokine receptor binding. This functional annotation concurs with the chromatin and gene regulation aspect of lncRNAs. Another significant correlation involves the fact that TERRA's downregulation impairs the TSS G4 structures associated with genes related to angiogenesis and cell division, including MYC and VEGFA. This decrease is associated with the modulation of cellular RNA levels, thereby further supporting the view that TERRA acts as a G4-stabilizing factor influencing gene activity. Other studies also have shown that lncRNAs such as Fendrr and MEG3 bind to chromatin modifiers to exert similar gene expression control. For instance, Fendrr mediates H3K4me3 and H3K27me3 marks through TrxG and PRC2 respectively. The binding of ATRX at G4 sites is reduced upon reduced TERRA levels, which causes destabilization of the promoter region for genes involved in cellular functions leading to its carcinogenesis. With reduced expression, another role for TERRA, like other lncRNAs, is in chromatin interaction and gene regulation. The use of epigenetic drugs such as hydralazine and magnesium valproate in mental illnesses certainly plays a role in the future of drug-based treatments for mental health, improving disease mitigation. The use of the various drugs mentioned in the treatment for major depressive disorders as explored in Candelaria (2007) is more effective in the presence of epigenetic assistance molecules in collaboration with traditional antidepressants, which would reduce the many risks for long-term use of antidepressants by reducing the dosage and term of the prescription. Together, this could mean the reduction of agitation, shakiness or anxious feelings, sick indigestion, stomach aches diarrhea, constipation loss of appetite, dizziness, insomnia, or feeling very sleepy headaches, loss of libido, erectile dysfunction, which are all imminent threats of the long term and high dosage use of antidepressants [(Side effects - Antidepressants, 2021)](). Together, these publications provide full insight into the interplay of epigenetics and ncRNAs in gene regulation, cellular homeostasis, and the progression of diseases.

    Not only are human diseases important in the grand scheme of the universe, but so are other organisms, such as plants and prokaryotic organisms. First, while the examples given in this paper reference the downsides of epigenetic factors' effects on wildlife (heavy metals causing epigenetic retardation of plant growth mentioned in Review 6), the mechanism at play between a wide variety of epigenetic factors that could include other non-harmful factors, such as aluminum sulfate or calcium oxide, which are both non-toxic to humans and perform magic in the garden, that could easily be turned into something that benefits the practical world, such as market sales, ecological preservation, agricultural sciences. To provide some enlightening examples, the commonly known example of hydrangeas changing color due to the acidity of its solid affecting its aluminum absorption levels, or any other cost-effective epigenetic modifications the industry can make to plants, could be used to produce favorable products for the market and be placed within the context of industry and the market. One way that epigenetics can be utilized in ecological preservation as a way to increase the resistance against stressors in many types of plants, such as Oryza sativa, researchers from Review 6 thoroughly studied over a long period with many many samples, the epigenetic traits of stress resistance heretic may lead to a burgeoning of wildlife that depend on it, leading to ecological success. Another example would be the use of cadmium-induced epigenetic changes (inducing DNA methylation in the hyperaccumulator may contribute to DNA protection and higher metal tolerance, also from Review 6) to make it easier for farmers and agricultural scientists to produce foodstuffs in contaminated soils, helping out places of less favorable agricultural land, often also correlated with places with less food security, to provide those in need with a possibility of growing crops in the ever-changing environment due to climate change and worldwide pollution.

    Prokaryotes are also a worthy candidate for discussion for new medicinal uses or research purposes. The current discoveries of epigenetic mechanisms have a lot of credit to give to prokaryotic organisms, which were easier to observe the use of drugs or other chemicals' effects. For example, a lot of the currently identified and thoroughly understood DNA methyltransferases were extensively studied on bacteria, which gave researchers the grounds to observe those drug's effects on

eukaryotes, such as the hairy roots of *S. miltiorhiza* Yang et al. (2022) to increase the production of secondary metabolites, or on rice, which possibly allows for the induction of acquired traits to be passed on to future generations, breaking the norm of Darwinian evolution (although to be fair, Darwin could have never accounted for man-made evolutionary genetical engineering).

Such insights should, in future studies, be harnessed into the development of targeted epigenetic treatments and diagnostic tools. These integrations will therefore be imperative to unravel the potential that ncRNAs, DNMTs, DNMTis, and other epi-drugs can provide in precision medicine, leading to better clinical outcomes and personalized treatment strategies against the plethora of diseases.

## CONCLUSION

The main method of organization used in this review paper was through the discussion of various papers detailing breakthrough discoveries. Each section consists of a contextualization that sets up the specific problem that is being solved, a summary of those research articles, connections between the various research mentioned, and a discussion of future implications and direction of epigenetic research. Finally, a discussion of those research articles was articulated to emphasize and highlight the potential uses and limitations of the research provided, as well as to reinforce and give alternative examples of the uses of research mentioned in the body. The intricate interplay between epigenetics and ncRNAs represents a complex regulatory network that significantly impacts gene expression, cellular homeostasis, and disease progression. Through mechanisms such as DNA methylation, histone modification, and RNA interference, ncRNAs play crucial roles in regulating gene expression and responding to environmental stimuli. This regulation is critical not only for understanding disease mechanisms but also for developing targeted therapies and diagnostic tools. In light of falling pieces about the molecular details of the regulation of epigenetic modification by ncRNA, the advent of ncRNA-based theranostics gets closer and closer. The development of these innovative points of use in personalized medicine may dramatically change the current concept of innovative treatments for different diseases including cancer and neurodegenerative diseases. One prominent example discussed in this paper is the role of ncRNAs in neuroblastoma, where aberrant DNA methylation and histone modifications are mediated by specific ncRNAs, leading to the dysregulation of oncogenes and tumor suppressor genes. The regulation of MYCN, one of the major oncogenes in neuroblastoma, by microRNAs (miRNAs) shows how the basic regulatory networks of cancer are explicitly intricate. Similarly, in cardiovascular diseases, it has been demonstrated that lncRNAs such as ANRIL control the chromatin state and gene expression implicating them in the pathogenesis of atherosclerosis. Furthermore, the involvement of ncRNAs in neurological diseases like Alzheimer's disease has potential, in which the dysregulated miRNAs are involved in the abnormal processing of amyloid precursor proteins and the worsening of disease conditions. The interaction of ncRNAs, especially miRNAs adds another layer of control in the context of Alzheimer's; so does epigenetic regulation. For example, particular miRNAs control the genes that contribute to amyloid-beta formation and tau protein aggregation. That is, it may be possible to change the course of the disease by targeting these miRNAs. These days there is a notion of "epigenetic therapy" and it starts showing the potential for the treatment of Alzheimer's. Treatments that can potentially help Alzheimer's disease include drugs like DNA methyltransferases (DNMTs) or histone deacetylase (HDACs). Conversely, similar clinical trials are being conducted for such conditions as chemoresistance, major depressive disorder, and chronic pain. This may prevent the defective expression of genes and neurophysiology while slowing down the progression of the respective ailments. The most important aspects of this review state that significant phenotypic alterations could arise from DNA methyltransferases and ncRNA in some organisms; particularly eukaryotes such as plants and animals besides prokaryotes like bacteria. There is a need for further studies to explore these epigenetic mechanisms to develop precision medicine approaches that offer personalized treatment strategies for various diseases since ncRNAs and epigenetic regulation are mutually interdependent in producing better healing effects. By harnessing the potential of ncRNAs and DNA methyltransferases (DNMTs), to create innovative therapies that address the root causes of gene expression abnormalities, as reflected in the clinical trials performed by Tiao-Hung Lai in MDD research.

Such advancements hold promise for improving clinical outcomes and enhancing the quality of life for patients with a range of genetic and epigenetic disorders, all under presumptively lower side effects than perhaps traditional antidepressants or invasive surgeries for other types of illnesses, such as chronic pain. Of particular interest in this regard is the use of SCARLET (Single-Cell Amplification and Real-Time Detection) sequencing discussed previously. This approach can give the quantitative measure of DNA methylation at the level of single cells, and thus, help with understanding the differences in epigenetic patterns between various cell subtypes and disease conditions. The capacity to dissect these patterns from such a fine level will certainly hold major significance on both the mechanisms by which epigenetic changes become associated with disease and on the identification of possible molecular targets, for therapeutic interventions. Further, recent work that employs the CRISPR-Cas9 system to regulate DNA methylation sites shows that epigenome editing might be a promising approach to treating diseases. Jia et.al showed that it is possible to intervene in precisely altering the methylation status of the genes so that normal expression patterns could be observed which may be useful in conditions such as cancer where DNA methylation is known to be involved abnormally. As well as for spinal muscular atrophy and to an extent, cancer, the major class of ncRNA antisense oligonucleotides have been shown to improve the recovery from the disease by mechanistically degrading or blocking the pathogenic ncRNAs produced by the diseases. The insights gained from studying epigenetic modifications in plants and prokaryotic organisms further underscore the universal importance of these mechanisms across different life forms. These studies not only provide a deeper understanding of basic biological processes but also highlight potential applications in agriculture, environmental conservation, and biotechnology, all of which benefit the quality of life for the average human on planet Earth. Overall, the field of epigenetics offers a promising frontier for scientific discovery and therapeutic innovation, with the potential to transform our approach to healthcare and disease management.

## CONFLICT OF INTEREST STATEMENT
The authors declare that the research was conducted in the absence of any commercial or financial interests that could be construed as a potential conflict of interest.

## EQUAL CONTRIBUTION STATEMENT
All authors declare that all authors contributed equally to this paper, and the order of the first author was decided through a coin toss.

## ACKNOWLEDGMENTS
We acknowledge the accreditation of the International Bilingual School at Hsinchu Science Park Taiwan affiliated with the National Experimental High School at Hsinchu Science Park Taiwan and associated administrative branches that allowed for the publishing of this paper.

## REFERENCES

*A Novel Mechanism of Long Non-Coding RNA in Epigenetic Regulation - spotlight - National Taiwan University*. (n.d.). National

   Taiwan University. https://www.ntu.edu.tw/english/spotlight/2022/2113_20221215.html

Akimoto, K. (2007, August). *Epigenetic Inheritance in Rice Plants*. Annals of Plant Biology. https://doi.org/10.1093/aob/mcm110

Benavides, M. P., Gallego, S. M., & Tomaro, M. L. (2005). Cadmium toxicity in plants. *Brazilian Journal of Plant Physiology*,

   *17*(1), 21–34. https://doi.org/10.1590/s1677-04202005000100003

Bilichek, A., Ilnytskyy, Y., Wóycicki, R., Kepeshchuk, N., Fogen, D., & Kovalchuk, I. (2015). The elucidation of stress memory inheritance in Brassica rapa plants. *Frontiers in Plant Science*, 6. https://doi.org/10.3389/fpls.2015.00005

Candelaria, M., Gallardo-Rincón, D., Arce, C., Cetina, L., Aguilar-Ponce, J., Arrieta, Ó., González-Fierro, A., Chávez-Blanco, A., De La Cruz-Hernández, E., Camargo, M., Trejo-Becerril, C., Pérez-Cárdenas, E., Pérez-Plasencia, C., Taja-Chayeb, L., Wegman-Ostrosky, T., Revilla-Vazquez, A., & Dueñas-González, A. (2007). A phase II study of epigenetic therapy with hydralazine and magnesium valproate to overcome chemotherapy resistance in refractory solid tumors. *Annals of Oncology*, *18*(9), 1529–1538. https://doi.org/10.1093/annonc/mdm204

Chmielowska-Bąk, J., Searle, I. R., Wakai, T. N., & Arasimowicz-Jelonek, M. (2023). The role of epigenetic and epitranscriptomic modifications in plants exposed to non-essential metals. *Frontiers in Plant Science*, *14*. https://doi.org/10.3389/fpls.2023.1278185

Collins, L. J., Schönfeld, B., & Chen, X. S. (2011). The epigenetics of non-coding RNA. In *Elsevier eBooks* (pp. 49–61). https://doi.org/10.1016/b978-0-12-375709-8.00004-6

Dai, Q., Ye, C., Irkliyenko, I., Wang, Y., Sun, H., Gao, Y., Liu, Y., Beadell, A., Perea, J., Goel, A., & He, C. (2024). Ultrafast bisulfite sequencing detection of 5-methylcytosine in DNA and RNA. *Nature Biotechnology*. https://doi.org/10.1038/s41587-023-02034-w

Duclot, F., & Kabbaj, M. (2015). Epigenetic mechanisms underlying the role of brain-derived neurotrophic factor in depression and response to antidepressants. *Journal of Experimental Biology*, *218*(1), 21–31. https://doi.org/10.1242/jeb.107086

Espinoza, C. A., Goodrich, J. A., & Kugel, J. F. (2007). Characterization of the structure, function, and mechanism of B2 RNA, an ncRNA repressor of RNA polymerase II transcription. *RNA*, *13*(4), 583–596. https://doi.org/10.1261/rna.310307

Fehér, A. (2015). Somatic embryogenesis — Stress-induced remodeling of plant cell fate. *Biochimica Et Biophysica Acta (BBA) - Gene Regulatory Mechanisms*, *1849*(4), 385–402. https://doi.org/10.1016/j.bbagrm.2014.07.005

Ferreira, H. J., & Esteller, M. (2018). Non-coding RNAs, epigenetics, and cancer: tying it all together. *Cancer and Metastasis Reviews*, *37*(1), 55–73. https://doi.org/10.1007/s10555-017-9715-8

Gaskin, D. J., & Richard, P. (2012). The economic costs of pain in the United States. *Journal of Pain*, *13*(8), 715–724. https://doi.org/10.1016/j.jpain.2012.03.009

Ghosh, D., Veeraraghavan, B., Elangovan, R., & Vivekanandan, P. (2020, January 27). Antibiotic resistance and epigenetics: More to it than meets the eye. Antimicrobial agents and chemotherapy. https://doi.org/10.1128/aac.02225-19

Gnyszka A, Jastrzebski Z, Flis S. DNA methyltransferase inhibitors and their emerging role in epigenetic therapy of cancer. Anticancer Res. 2013 Aug;33(8):2989-96.

Hassan, M., Amir, A., Shahzadi, S., & Kloczkowski, A. (2022, November 4). *Therapeutic implications of microRNAs in depressive disorders: A Review*. MDPI. https://doi.org/10.3390/ijms232113530

Huang, T.-L., & Lin, C.-C. (2015). Advances in biomarkers of major depressive disorder. *Advances in Clinical Chemistry*, 177–204. https://doi.org/10.1016/bs.acc.2014.11.003

Jacob, R., Zander, S., & Gutschner, T. (2017, November 10). *The Dark Side of the epitranscriptome: Chemical Modifications in long non-coding RNAs*. MDPI. https://doi.org/10.3390/ijms18112387

Jin, B., & Robertson, K. D. (2012). DNA methyltransferases, DNA damage repair, and cancer. *Advances in Experimental Medicine and Biology*, 3–29. https://doi.org/10.1007/978-1-4419-9967-2_1

Kaspar, D., Hastreiter, S., Irmler, M., De Angelis, M. H., & Beckers, J. (2020). Nutrition and its role in the epigenetic inheritance of obesity and diabetes across generations. *Mammalian Genome*, *31*(5–6), 119–133. https://doi.org/10.1007/s00335-020-09839-z

Kazimierczyk, M., & Wrzesinski, J. (2021, June 7). *Long non-coding RNA epigenetics*. MDPI. https://doi.org/10.3390/ijms22116166

Kiran, N., Bharti, R., & Sharma, R. (2022). Effect of heavy metals: An overview. *Materials Today Proceedings*, *51*, 880–885. https://doi.org/10.1016/j.matpr.2021.06.278

Lamka, G. F., Harder, A. M., Sundaram, M., Schwartz, T. S., Christie, M. R., DeWoody, J. A., & Willoughby, J. R. (2022, March 16). *Epigenetics in ecology, evolution, and conservation*. Frontiers. https://doi.org/10.3389/fevo.2022.871791

Li, Y., & Tollefsbol, T. O. (2011). *DNA methylation detection: Bisulfite genomic sequencing analysis*. Methods in molecular biology (Clifton, N.J.). https://doi.org/10.1007/978-1-61779-316-5_2

Lillycrop, K. A., & Burdge, G. C. (2012). Epigenetic mechanisms linking early nutrition to long-term health. *Best Practice & Research Clinical Endocrinology & Metabolism*, *26*(5), 667–676. https://doi.org/10.1016/j.beem.2012.03.009

Mangiavacchi, A., Morelli, G., & Orlando, V. (2023). Behind the scenes: How RNA orchestrates the epigenetic regulation of gene expression. *Frontiers in Cell and Developmental Biology*, *11*. https://doi.org/10.3389/fcell.2023.1123975

Mauceri, D. (2022). Role of epigenetic mechanisms in chronic pain. *Cells*, *11*(16), 2613. https://doi.org/10.3390/cells11162613

Mercer, T. R., Dinger, M. E., & Mattick, J. S. (2009). Long non-coding RNAs: insights into functions. *Nature Reviews Genetics*, *10*(3), 155–159. https://doi.org/10.1038/nrg2521

Nestler, E. J., Peña, C. J., Kundakovic, M., Mitchell, A., & Akbarian, S. (2016). Epigenetic basis of mental illness. *The Neuroscientist*, *22*(5), 447–463. https://doi.org/10.1177/1073858415608147

Sathipati, S. Y., Sahu, D., Huang, H., Lin, Y., & Ho, S. (2019). Identification and characterization of the lncRNA signature associated with overall survival in patients with neuroblastoma. *Scientific Reports*, *9*(1). https://doi.org/10.1038/s41598-019-41553-y

Shafik, A., Schumann, U., Evers, M., Sibbritt, T., & Preiss, T. (2016). The emerging epitranscriptomics of long noncoding RNAs.

*Biochimica Et Biophysica Acta (BBA) - Gene Regulatory Mechanisms*, *1859*(1), 59–70.

https://doi.org/10.1016/j.bbagrm.2015.10.019

*Side effects - Antidepressants*. (2021, November 18). nhs.uk.

https://www.nhs.uk/mental-health/talking-therapies-medicine-treatments/medicines-and-psychiatry/antidepressants/side-effects/

Statello, L., Guo, C., Chen, L., & Huarte, M. (2020). Gene regulation by long non-coding RNAs and their biological functions. *Nature Reviews Molecular Cell Biology*, *22*(2), 96–118. https://doi.org/10.1038/s41580-020-00315-9

Szcześniak, M. W., & Makałowska, I. (2016). lncRNA-RNA Interactions across the Human Transcriptome. *PLoS ONE*, *11*(3), e0150353. https://doi.org/10.1371/journal.pone.0150353

Tsai, R., Fang, K., Yang, P., Hsieh, Y., Chiang, I., Chen, Y., Lee, H., Lee, J. T., & Chu, H. C. (2022). TERRA regulates DNA G-quadruplex formation and ATRX recruitment to chromatin. *Nucleic Acids Research*, *50*(21), 12217–12234. https://doi.org/10.1093/nar/gkac1114

Wang, C., Wang, L., Ding, Y., Lu, X., Zhang, G., Yang, J., Zheng, H., Wang, H., Jiang, Y., & Xu, L. (2017). LNCRNA structural characteristics in Epigenetic regulation. *International Journal of Molecular Sciences*, *18*(12), 2659. https://doi.org/10.3390/ijms18122659

Wang, X., Yu, D., & Chen, L. (2023, June 14). *Antimicrobial resistance and mechanisms of epigenetic regulation*. Frontiers in cellular and infection microbiology. https://doi.org/10.3389/fcimb.2023.1199646

Wei, J., Huang, K., Yang, C., & Kang, C. (2016). Non-coding RNAs as regulators in epigenetics. Oncology Reports, 37(1), 3–9. https://doi.org/10.3892/or.2016.5236

*What is Epigenetics? The Answer to the Nature vs. Nurture Debate*. (2020, October 30). Center on the Developing Child at Harvard University.

https://developingchild.harvard.edu/resources/what-is-epigenetics-and-how-does-it-relate-to-child-development/#:~:text=%E2%80%9CEpigenetics%E2%80%9D%20is%20an%20emerging%20area,in%20stone%E2%80%9D%20has%20been%20disproven

Erika Wolff, Gangning Liang, Yvonne C. Tsai, J. Esteban Castelao, Xuejuan Jiang, Susan L. Groshen, Peter A. Jones; Association of hypermethylation with grade, stage, and years of tobacco smoking in transitional cell carcinoma. Cancer Res 15 April 2006; 66 (8_Supplement): 10–11.

Varela, M. A., Roberts, T. C., & Wood, M. J. A. (2013, September 26). *Epigenetics and ncRNAs in brain function and disease: Mechanisms and prospects for therapy - neurotherapeutics*. SpringerLink. https://doi.org/10.1007/s13311-013-0212-7

Yang, B., Lee, M., Lin, M., & Chang, W. (2022). 5-Azacytidine increases tanshinone production in Salvia miltiorrhiza hairy roots through epigenetic modulation. *Scientific Reports*, *12*(1). https://doi.org/10.1038/s41598-022-12577-8

Zhao, J., Sun, B. K., Erwin, J. A., Song, J., & Lee, J. T. (2008). Polycomb proteins are targeted by a short repeat RNA to the mouse X chromosome. *Science*, *322*(5902), 750–756. https://doi.org/10.1126/science.1163045